\pdfoutput=1
\documentclass[letterpaper,twocolumn,10pt]{article}
\usepackage{usenix-2020-09}

\usepackage{tikz}
\usepackage{amsmath}
\usepackage{graphicx}
\usepackage{subcaption}
\usepackage{soul}
\usepackage[normalem]{ulem}
\usepackage{booktabs}
\usepackage{algorithm}
\usepackage{algorithmic}
\usepackage{amsmath}
\usepackage{multirow}
\usepackage{xurl}
\usepackage{pifont}

\usepackage[compact]{titlesec}
\newcommand{\tightcaption}[1]{ \vspace{-0.2cm}\caption{\small{#1}}\vspace{-0.2cm}}

\newcommand*\whitecircled[1]{%
  \tikz[baseline=(char.base)]{%
    \node[shape=circle,draw=black,fill=white,inner sep=1.5pt] (char)
         {\color{black}\footnotesize #1};}}
\newenvironment{packeditemize}{
\begin{list}{$\bullet$}{
\setlength{\itemsep}{1.5pt}
\setlength{\labelwidth}{8pt}
\setlength{\leftmargin}{10pt}
\setlength{\labelsep}{3pt}
\setlength{\listparindent}{\parindent}
\setlength{\parsep}{1.5pt}
\setlength{\parskip}{1.5pt}
\setlength{\topsep}{1.5pt}}}{\end{list}}

\expandafter\def\expandafter\normalsize\expandafter{%
    \normalsize%
    \setlength\abovedisplayskip{2pt}%
    \setlength\belowdisplayskip{1pt}%
    \setlength\abovedisplayshortskip{0pt}%
    \setlength\belowdisplayshortskip{0pt}%
}

\usepackage{filecontents}

\usepackage[available,functional,reproduced]{usenixbadges}

\begin{document}

\pagestyle{empty}

\date{}

\title{\Large \bf R2CC: Reliable and Resilient Collective Communication for Distributed Learning}

\author{
{\rm Wei Wang, Nengneng Yu, Sixian Xiong, Zaoxing Liu}\\
University of Maryland
}

\maketitle

\begin{abstract}
Modern ML training and inference now span hundreds to tens of thousands of GPUs, where failures can waste 10–15\% of GPU hours due to slow recovery. Common network errors and link fluctuations can cause up to 50\% of job interruptions, triggering timeouts that often terminate entire jobs, forcing expensive checkpoint rollback during training and request reprocessing during inference. We present R2CC, a fault-tolerant communication library that provides lossless, low-overhead failover by exploiting multi-NIC hardware. R2CC performs rapid connection migration, bandwidth-aware load redistribution, and resilient collective algorithms to maintain progress under failures. We evaluate R2CC on four 8-GPU H100 InfiniBand servers and via large-scale ML simulators modeling up to 1024 GPUs with diverse failure patterns. Experiments show that R2CC is highly robust to inter-node network failures, incurring less than 1.1\% training and 3\% inference overhead under active failure scenarios. Compared to existing fault-tolerant approaches, R2CC reduces failure-induced overhead by up to 92\% for training and 98\% for inference. R2CC is open-sourced at \url{https://github.com/r2cc-project/R2CC}.

\end{abstract}

\section{Introduction}

The recent success of large-scale machine learning models is fundamentally reshaping the computing landscape~\cite{hu2024characterization, choudhury2024mast, gangidi2024rdma, qian2024alibaba}. Training and serving large models require massive distributed GPU clusters: a fragile and resource intensive endeavor, highly susceptible to various hardware and software failures~\cite{megascale, gandhi2024recycle}. For example, pre-training of LLaMA3 on a cluster of 16,384 GPUs experienced 466 job interruptions during a 54-day training period~\cite{llama3} and the mean-time-to-failure was only 2.7 hours. Once a failure occurs, a noticeable recovery time is often required~\cite{unicron,megascale}.

While network communication is one of the key performance bottlenecks in ML training~\cite{Megascalemoe,zhou2025accelerating,wang2023zeroplusplus,song2023optimus}, network failures and interruptions remain an under-addressed source of inefficiency~\cite{wang2024towards, yu2023understanding, Revisiting_Reliability}.
Recent industry reports show that network issues are among the leading causes of job failures in large ML clusters~\cite{llama3, Revisiting_Reliability, liu2024hostmesh, wan2025robust}, and faults such as cable and NIC issues can account for up to half of all failed training jobs~\cite{unicron}. Similarly, these failures also affect inference systems, where even a single stalled collective can significantly increase tail latency or trigger costly request retries~\cite{ICML2024}. A key reason is that current collective communication libraries (CCLs), such as NCCL~\cite{NCCL}, are {\em fail-stop}: any network error aborts the communicator, crashes the process, and forces the entire job to restart.

To recover from failures, existing ML frameworks fall into one of two extremes: (1) A checkpointing-based recovery process requires a \textit{complete relaunch} of the framework to restore the model's state from storage, incurring substantial latency for failure detection, resource re-allocation, state reloading, and re-computation of iterations lost since the last checkpoint~\cite{bytecheckpoint,eisenman2022check,gupta2024just,mohan2021checkfreq,wang2023gemini}. Recent efforts~\cite{unicron,megascale} show that the recovery time from a single failure can reach 68 minutes, wasting large amounts of GPU time in jobs~\cite{wang2023gemini}. (2) An emerging response is to exclude the failed server node~\cite{oobleck,gandhi2024recycle}. When any failure occurs, the framework runtime removes that worker along with its GPUs, rebuilds the job, and communicates over the remaining ranks, thereby avoiding a full restart while still interrupting progress and reducing the effective usage of GPUs if only network issues are present.

From a systems layering perspective, these approaches are unsatisfactory because they handle network failures at the application (ML framework) layer rather than within the communication layer. Analogous to how transport protocols such as TCP provide reliability for packet loss~\cite{saltzer1984end,zimmermann1980osi}, the CCL layer should itself offer a more \textit{resilient} abstraction under network faults. In other words, there ought to be a \textit{middle ground} in which the CCL transparently routes around transient link and NIC failures, degrades gracefully, and exposes meaningful signals to guide higher-level recovery decisions.

One opportunity to improve CCL resilience in modern multi-NIC GPU clusters is the increasing heterogeneity of network link types between GPUs (e.g., PCIe and NVLink), which creates potentially redundant network paths for communication, even if some of these paths may fail. In this paper, we want to explore whether there are alternative cases in which
\textbf{\textit{CCL layer can exploit the path redundancy to continue the ongoing job efficiently rather than giving up immediately.}}

\begin{figure}[t]
    \centering
    \includegraphics[width=0.95\columnwidth]{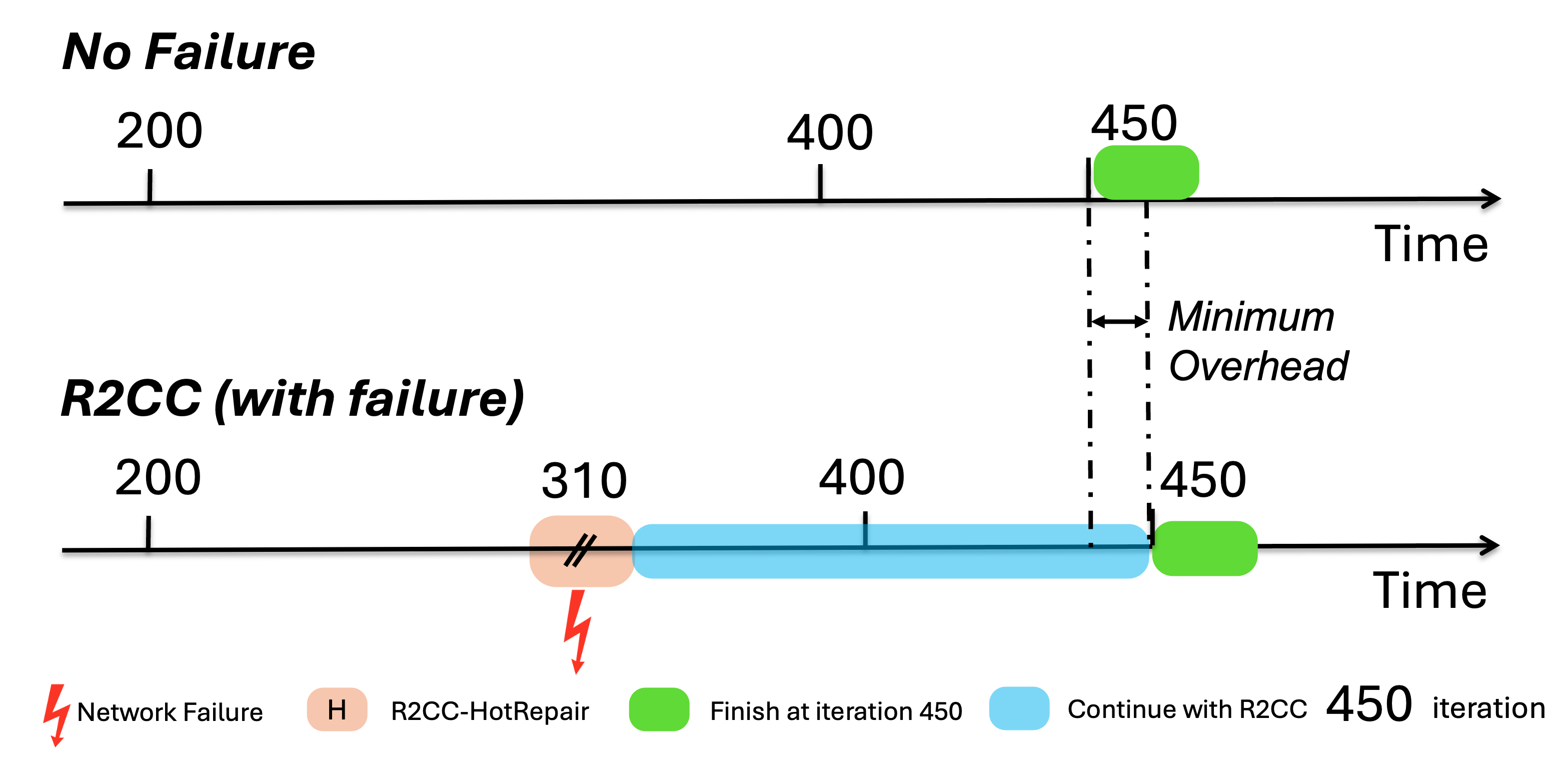}
    \vspace{-2mm}
\tightcaption{\textbf{R2CC (with failure) vs. Ideal (failure-free)}}
    \label{fig:r2ccandnofailure}
    \vspace{-2mm}
\end{figure}

In this paper, we present R2CC, a reliable and resilient collective communication library that transforms collective communication from fail-stop to fail-operational under inter-node network failures (Figure~\ref{fig:r2ccandnofailure}). When a NIC or link fails but alternative paths remain available, R2CC detects faults during ongoing collectives, migrates in-flight transfers losslessly to healthy connections, and adapts collective schedules to the degraded topology. As a drop-in extension for NCCL~\cite{NCCL}, R2CC requires no framework changes for transparent repair and optimization. R2CC additionally exposes degradation signals, reporting failure location and bandwidth loss, enabling frameworks that integrate with R2CC to make informed policy decisions on continuation, checkpointing, or recovery. To realize this, we address the following key challenges.

{\bf \#1: How to detect and locate the failure in an ongoing collective operation (e.g., AllReduce)?} Detecting and pinpointing failures during an ongoing collective is difficult due to limited error visibility in RDMA. R2CC introduces bilateral failure awareness using an out-of-band notification path combined with probe-based RDMA fault localization.

{\bf \#2: How to losslessly migrate the failed connection to another connection via alternative links?} R2CC performs live migration to alternative paths using two techniques. First, \textit{eager provisioning} multi-registers GPU buffers with local NICs and pre-establishes backup RDMA connections to neighboring servers at initialization. Second, \textit{protocol-aware suffix replay} reconciles a conservative replay boundary with the peer over the out-of-band channel, so the collective can safely replay only the unfinished suffix on the backup connection under both throughput-oriented and low-latency CCL protocols.
These techniques allow collectives to resume seamlessly on a backup NIC.

{\bf \#3: How to perform failure-aware collective online optimization  with single or multiple failures?} R2CC selects from several failure-aware optimizations. \textit{R2CC-Balance} redistributes traffic across remaining NICs (PCIe-, NUMA-, and PXN-aware) for all collectives. \textit{R2CC-AllReduce} modifies the AllReduce schedule to reduce the workload on the degraded server, combining global and partial AllReduce operations. Third, for multi-failure scenarios, an extended topology-aware \textit{recursive R2CC-AllReduce} can navigate bandwidth asymmetry and rail mismatches.

We evaluate R2CC on a testbed consisting of four NVIDIA DGX servers, each equipped with 8$\times$H100 GPUs and 8 InfiniBand 400Gbps NICs. We complement this with large-scale simulations using the state-of-the-art training simulator SimAI~\cite{wang2025simai} to model clusters of up to 1024 GPUs under diverse failure patterns.
With a single NIC failure injected (i.e., a 12.5\% bandwidth loss in the affected server), our experiments show that R2CC incurs less than 1.1\% overhead for training and less than 3\% for inference. Compared with recent fault-tolerant systems for training and inference, R2CC achieves up to 92\% and up to 98\% lower overheads respectively.

\section{Background and Motivation}

In this section, we discuss the hardware and software stacks in ML training and serving platforms. We explore the possible failure handling solutions and examine why they can be ill-suited to address network failures.

\subsection{Hardware Stack and Common Failures }
\label{sec:bg-stack}

\noindent\textbf{Hardware stack in GPU Clusters.}
Modern GPU clusters employ a two-tier communication fabric that separates high-bandwidth intra-node communication from constrained inter-node links.\footnote{Examples include production LLM training clusters reported by Meta~\cite{gangidi2024rdma}, Alibaba~\cite{qian2024alibaba}, and others.}
Within a server node, GPUs are connected via NVLink or NVSwitch, providing hundreds of GB/s of bidirectional bandwidth and low latency between devices on the same node.
Between nodes, GPUs communicate through RDMA-capable NICs (e.g., InfiniBand or RoCE) attached to the host over PCIe.
Because a single NIC typically offers substantially less bandwidth than the intra-node fabric, production servers aggregate multiple NIC ``rails'' per node, and the effective inter-node connection between a pair of servers consists of several physical paths that may traverse different switch ports or even different ToR switches.
Recent designs, such as multi-plane fabrics~\cite{qian2024alibaba,oci-multiplanar,khashab2026high} and GPU servers with directly attached NICs, change how NICs connect to switches and GPUs; Appendix~\ref{app:discussion} discusses R2CC under these designs.
GPU buffers are exposed to these NICs via memory registration, allowing NICs to perform direct RDMA reads and writes on GPU memory without CPU-side copies, but only along links for which registration and queue pairs have been established.

\smallskip\noindent\textbf{Common hardware failures.}
Failures in this stack fall into three categories: GPU failures, intra-node link failures, and inter-node network failures. GPU failures include hardware, memory, or other errors that render a device unusable. Intra-node link failures are faults in the GPU fabric within a server. Inter-node network failures are faults that affect the available bandwidth or the set of network paths between two otherwise live GPU endpoints.

\smallskip\noindent\textbf{R2CC scope.}
As a CCL, R2CC targets \textbf{inter-node network failures} that change the available paths or bandwidth between live GPU endpoints, while at least one alternate path remains usable. This includes hard faults such as NIC, port, cable, or ToR-port failures, as well as gray failures, such as link flapping, that surface as transport errors. GPU failures that remove the compute endpoint, including rare cases where network faults co-occur with GPU failures~\cite{Revisiting_Reliability}, are out of scope; they are typically handled by training and inference frameworks. Intra-node link failures are also out of scope because proprietary protocols such as NVLink/NVSwitch lack a CPU/CCL-visible recoverable transport boundary, such as per-transfer completion/error reporting and replay state. Supporting such failures would require redesigning the intra-node transport protocol. Intra-node link failures are also not reported as frequent errors in the hardware stack. Additional discussion is provided in Appendix~\ref{app:failure-scope}.

\subsection{Software Stack}
\label{sec:bg-failures}
\begin{figure}[t]
    \centering
    \includegraphics[width=0.97\columnwidth]{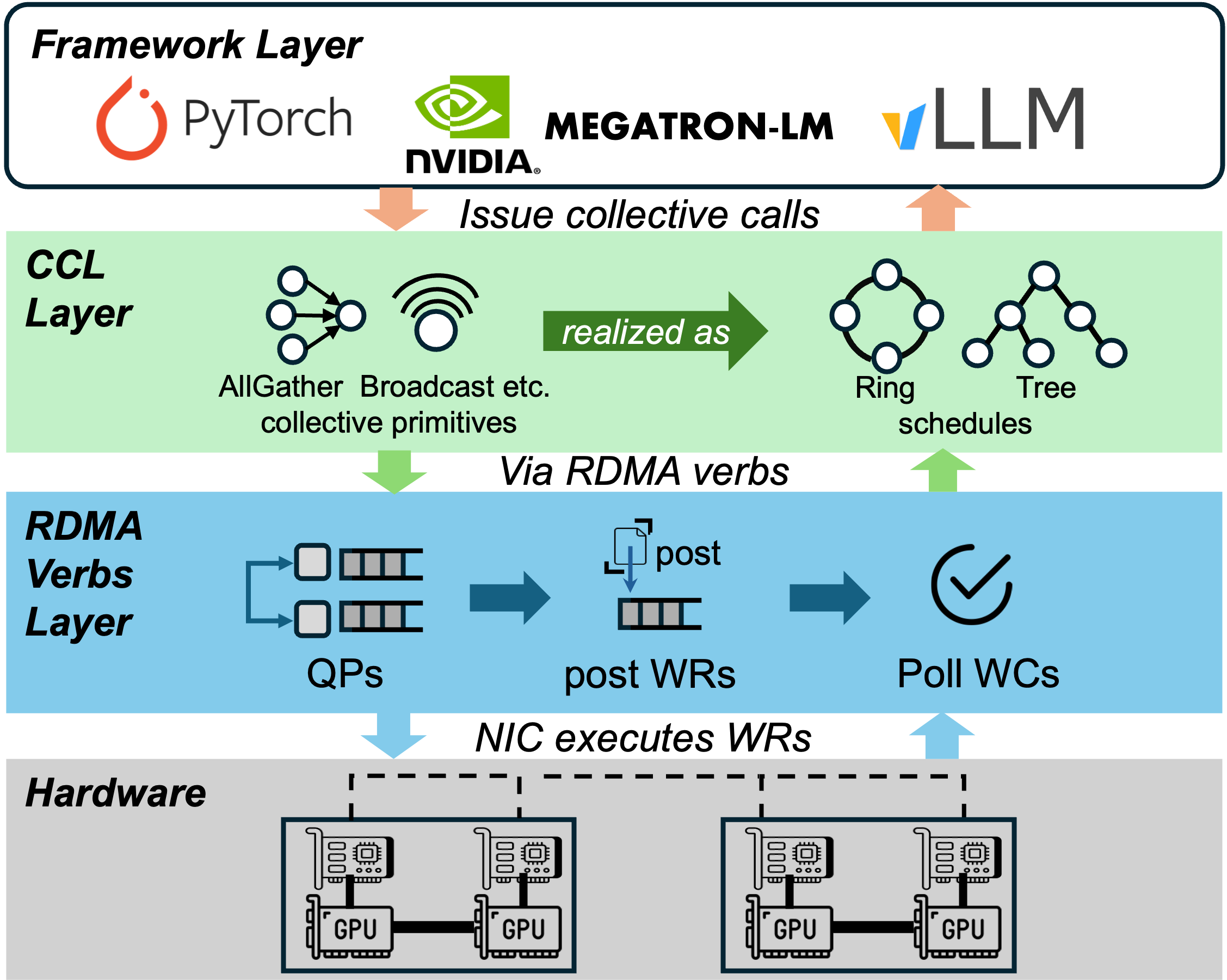}
    \tightcaption{\textbf{Software layering of collective communication.}
    }
    \label{fig:layer}
    \vspace{-3mm}
\end{figure}

Figure~\ref{fig:layer} illustrates the software layering of distributed collective communication.

\smallskip\noindent\textbf{Framework layer.}
The ML framework layer owns the training or inference loop. Systems such as PyTorch~\cite{paszke2019pytorch}, DeepSpeed~\cite{rasley2020deepspeed}, and Megatron~\cite{shoeybi2019megatron} partition work across GPUs and maintain model or request state across iterations. They drive training steps such as forward, backward, and optimizer updates, or serving request steps such as prefilling and decoding~\cite{kwon2023efficient,agrawal2024taming,zhong2024distserve}. The framework delegates cross-GPU data movement to the CCL layer through collective or point-to-point operations.

\smallskip\noindent\textbf{CCL layer.}
The CCL layer implements the collective and point-to-point operations delegated from the framework. Libraries such as NCCL~\cite{NCCL}, RCCL~\cite{RCCL}, and UCCL~\cite{zhou2025extensible} organize the participating GPUs into channels and realize each collective as a ring or tree schedule. In a ring schedule, data is partitioned into chunks that circulate across GPUs, advancing one hop per step. For collectives such as Broadcast, ReduceScatter, or AllGather with input size $D$, each server sends and receives approximately $D$ bytes. AllReduce is commonly implemented as a ReduceScatter followed by an AllGather, resulting in 2D bytes of communication per server.

Each transfer step in CCL follows a specific performance-oriented protocol. In throughput-oriented protocols, bulk data transfer is decoupled from completion notification: the GPU processes data only after the receiver’s CPU observes a corresponding notification. In contrast, low-latency protocols embed both payload and progress metadata within the same transfer, allowing the GPU to consume data incrementally as it arrives. The CCL layer coordinates these transfers and drives communication progress.

\smallskip\noindent\textbf{RDMA verbs layer.}
The RDMA verbs layer provides the interface between the CCL and the network hardware. Inter-node communication is realized through queue pairs (QPs) established between endpoints. The CCL issues work requests (WRs) to initiate communication and polls completion queues (CQs) for work completions (WCs), which indicate progress or errors. Two operations are commonly used here: RDMA Write transfers data directly into a pre-registered remote GPU buffer without involving the receiver's CPU. RDMA Write with Immediate (WriteImm) additionally carries a small inline value, consumes a pre-posted receive request on the remote side, and generates a completion event to notify the receiver. These primitives enable high-throughput, low-overhead communication but provide only limited visibility into higher-level communication semantics.

\subsection{Current Failure Handling and Limitations}
\label{sec:related}
\noindent\textbf{Framework-level failure handling.} The framework layer handles hardware failures at the granularity of GPU ranks or entire server nodes. Checkpoint-based approaches, such as Gemini~\cite{wang2023gemini} and ByteCheckpoint~\cite{bytecheckpoint}, provide fast in-memory restore, efficient save/load, and resharding across training configurations. Other systems, such as TrainMover~\cite{lao2024trainmover}, use standby machines and live migration to replace failed workers while preserving the training configuration. Oobleck~\cite{oobleck} and ReCycle~\cite{gandhi2024recycle} adapt training execution after worker changes through pipeline templates, replica recovery, or pipeline-schedule adaptation. On the model serving side, D\'{e}j\`{a}Vu~\cite{ICML2024} streams and replicates KV-cache state so that requests can resume on a healthy replica. All of these approaches implicitly assume that the compute endpoint itself is lost or must be removed. As a result, they recover by replacing or excluding entire workers. They are therefore ill-suited for failures where both GPU endpoints remain operational and only a subset of the inter-node network paths is affected.

\smallskip\noindent\textbf{RDMA transport-level failure handling.}
At the RDMA transport layer, failures manifest as queue pair (QP) errors or retry-exhausted operations. For instance, SHIFT~\cite{shift} and LubeRDMA~\cite{luberdma} attempt to mask these failures by transparently reissuing outstanding work requests over alternative paths, allowing transfers to complete without exposing errors to higher layers. However, these mechanisms operate at the level of work requests and completions, without visibility into collective schedules. Under modern low-latency CCL protocols (e.g., LL/LL128 in NCCL~\cite{NCCL}), GPUs may consume data before the corresponding completion is observed by the CPU, which is outside the transport layer's control. As a result, systems like SHIFT~\cite{shift} need to disable low-latency protocols to maintain data correctness at the cost of performance.

\smallskip\noindent\textbf{CCL-layer responses.} At the collective communication layer, mainstream libraries such as NCCL and RCCL have long adopted fail-stop semantics: any transport error aborts the collective, terminates the process, and forces job-level recovery. Although the latest NCCL versions (v2.29.7 and later) support active-active failover across multiple ports or NICs assigned to a GPU, this approach reduces bandwidth as it fails to rebalance collective workloads and schedules. 
Other recent efforts attempt to improve resilience within the CCL layer. For example, AdapCC~\cite{zhao2024adapcc} detects non-responsive workers via timeouts and coordinates with an elastic training controller to continue with a reduced worker set, but it still fails once a fatal error occurs during an ongoing collective. ICCL~\cite{chen2025efficient} pre-establishes primary and backup connections between GPU pairs and migrates traffic upon failure; however, it routes all traffic through a single backup path without adjusting the collective schedule, creating a bottleneck on the affected node.

We observe that \textbf{the key missing piece in today's software stack is a resilient CCL layer.}
Inter-node network failures should be handled within the CCL layer rather than escalating to process or job-level failures. Ideally, the CCL should continue ongoing collectives transparently and correctly, so that collective operations remain indistinguishable from failure-free execution. Beyond merely preserving correctness, it should also adapt communication schedules to the degraded topology and mitigate performance bottlenecks caused by reduced bandwidth. Finally, it should expose degradation signals to upper layers, enabling frameworks to make informed decisions about continuation or recovery, rather than reacting only after a crash occurs.

\section{R2CC Overview}

\begin{figure}
    \centering
    \includegraphics[width=1\columnwidth]{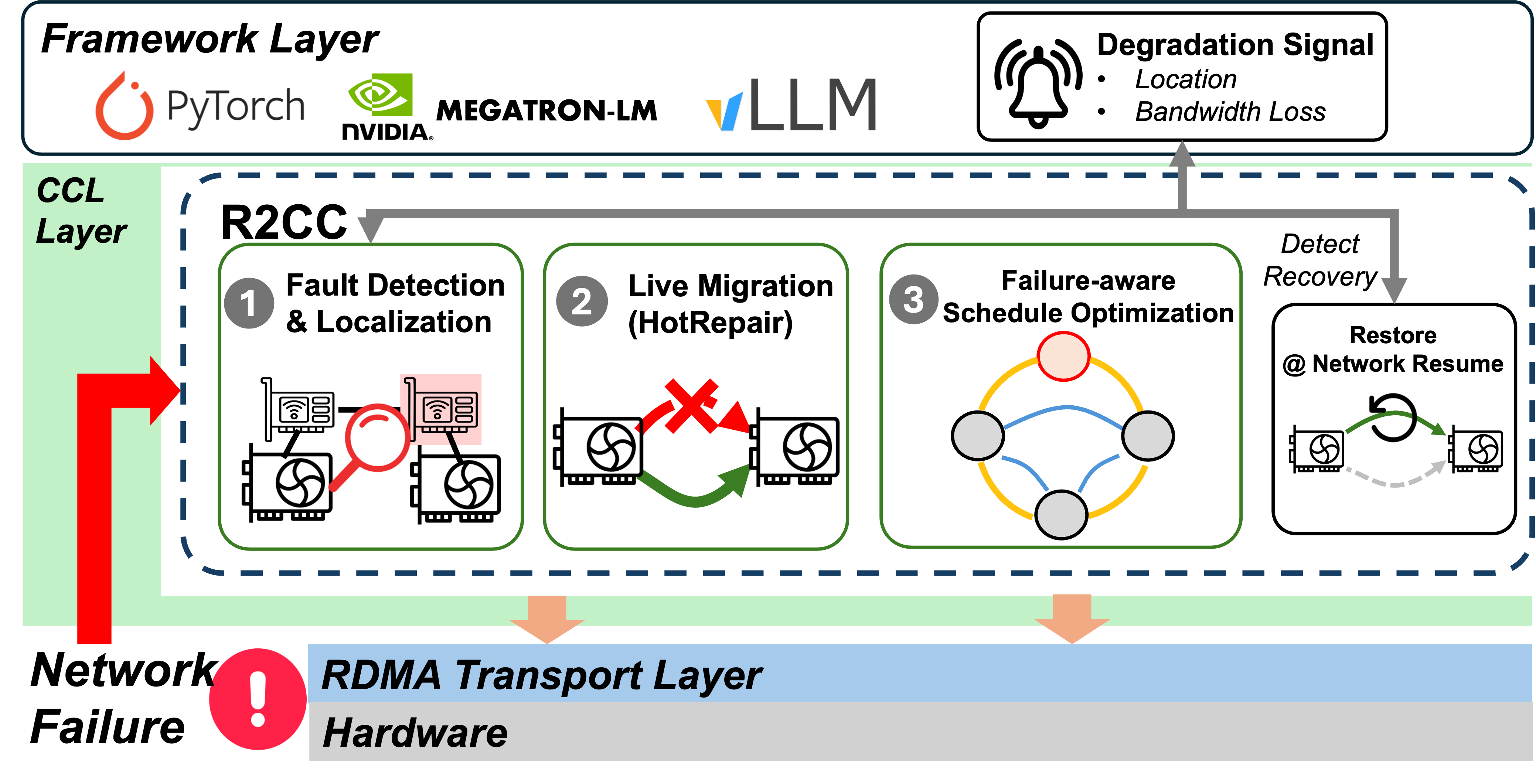}
    \vspace{-15pt}
    \tightcaption{\textbf{R2CC Overview.}}
    \label{fig:architecture}
    \vspace{-2mm}
\end{figure}

Figure~\ref{fig:architecture} summarizes how R2CC handles network failures during collective communication. R2CC is implemented as a drop-in extension of NCCL with RDMA support, the most widely deployed CCL for ML training and inference. When an inter-node failure occurs, R2CC transforms the CCL from fail-stop to fail-operational through three key components:

\smallskip\noindent\whitecircled{1}~\textbf{Fault detection and localization.} When a connection error surfaces during a collective (e.g., an immediate QP error, RDMA retry exhaustion), R2CC intercepts it and turns it into a recoverable CCL-managed event, then quickly localizes the fault to the affected side (local, remote, or bilateral). The localization result then drives live migration onto a backup connection (\S\ref{sec:design-mig}) and failure-aware scheduling (\S\ref{sec:single-failure}).

\smallskip\noindent\whitecircled{2}~\textbf{Live migration (HotRepair).} R2CC then migrates the ongoing collective onto a pre-established backup RDMA connection. Over the out-of-band (OOB) path, the two endpoints reconcile a replay boundary using NCCL protocol semantics rather than transport state alone (\S\ref{sec:design-mig}). This keeps replay correct even for CCL low latency protocol variants. Only the unreconciled suffix is retransmitted on the backup connection.

\smallskip\noindent\whitecircled{3}~\textbf{Failure-aware collective schedule optimization.}
HotRepair preserves correctness, but it does not, by itself, adapt the collective to the degraded heterogeneous topology. \textbf{R2CC-Balance} is the general optimization: it redistributes traffic from the failed NIC across surviving paths (e.g., PCIe forwarding or NVLink PXN\footnote{NVIDIA PXN~\cite{nccl-pxn} routes inter-server GPU traffic through intermediate GPUs as proxies.}) without changing the collective pattern, and applies broadly across collectives and P2P communication. For throughput-bound AllReduce on large messages, \textbf{R2CC-AllReduce} adds a specialized optimization on top of R2CC-Balance: it modifies the standard ring schedule to reduce data sent to and from the impaired server, further improving performance under degraded connectivity.

\smallskip\noindent\textbf{Degradation signaling and restore.}
As soon as a fault is localized, R2CC reports the degradation status to the framework, so the upper layer has both more policy choices and more time to act than it would under a later crash or abort. It can choose whether to continue running, run to its next safe boundary and then act, shrink, or restart. In the background, R2CC periodically probes the affected path. When the path becomes usable again, R2CC restores the original route, removes the corresponding scheduling adaptation, and updates the degradation signal.

\subsection{Challenges and Key Ideas}

\textbf{C1: Error visibility in an ongoing collective.} Hot repair must identify and pinpoint failures mid-collective before NCCL's default crash-on-error behavior triggers. In practice, NIC failures surface only as generic connection errors: the network stack cannot tell which endpoint actually failed, which creates \textit{asymmetric error visibility} that can leave peers waiting on a dead connection.

\textit{Key Idea:} R2CC introduces \textit{bilateral failure awareness} over a bootstrap network, combined with \textit{probe-based RDMA fault localization}, to rapidly locate the faulty endpoint. When a link between servers A and B fails, R2CC immediately notifies both sides over the out-of-band notification path to eliminate half-open states where one endpoint continues waiting on a dead connection, then triangulates the fault via three-point probing from a third peer in the same communicator.

\smallskip\noindent\textbf{C2: Lossless live migration over alternative links.} Even once the faulty NIC is known, NCCL's tight GPU--NIC binding prevents an ongoing collective from switching to an alternative path. Migration must also be lossless: the backup connection has to resume the collective without dropping or duplicating any in-flight chunk, under both throughput-oriented and low-latency CCL protocols.

\textit{Key Idea:} To break NCCL's GPU–NIC binding, R2CC pre-establishes idle backup connections to neighboring servers at initialization. These idle backup connections enable R2CC to reconcile a protocol-aware replay boundary with the peer and resume the collective on the backup connection without reinitializing NCCL.

\smallskip\noindent\textbf{C3: Online collective bandwidth optimization with single or multiple failures.} Network failures transform the cluster into a heterogeneous topology. Naively redirecting all traffic to a single backup NIC (i.e., basic HotRepair) often leads to congestion and bottlenecks. While solver-based schedule synthesis can find optimal routing, its high computational latency makes it unsuitable for real-time recovery. The primary challenge lies in rapidly generating failure-aware schedules that maximize bandwidth utilization within a degraded network.

\textit{Key Idea:} R2CC-Balance treats NICs as a unified resource pool, redistributing traffic through topology-aware heuristics. R2CC-AllReduce adapts the standard collective algorithm by proportionally balancing data loads according to degraded bandwidth. To address complex concurrent failures, R2CC generalizes these mechanisms through topology-aware logical re-ranking and recursive R2CC-AllReduce scheduling, maintaining high throughput across increasingly degraded topologies.

\section{Failure Detection and Mitigation}\label{sec:design-mig}

When network errors occur during active communication, CCLs typically trigger crashes that disrupt the entire ML job. R2CC intercepts these errors before they escalate into fatal crashes. In this section, we discuss R2CC's real-time failure detection and mitigation mechanism that consists of
(1) \textit{bilateral failure awareness} to quickly determine that a problem exists, (2) \textit{precise fault localization} to identify where the failure is, and (3) \textit{live migration} to continue the communication with alternative connections. We also introduce \textit{degradation signaling and restore} to detect recovered paths and restore the original route.

\subsection{Bilateral Error Awareness}

When a network error occurs, the server node that detects the fault typically crashes, while others in the collective remain unaware and continue polling for progress that will never occur. This asymmetric failure visibility stems from collective frameworks like NCCL, whose asynchronous, pipelined backends (e.g., RDMA, TCP) provide no fine-grained, in-band failure signaling. RDMA in this case is especially problematic: NICs autonomously handle retries and one-sided operations, delaying error propagation to the CPU.

\smallskip\textbf{Our approach:} R2CC provides bilateral failure awareness through an out-of-band (OOB) notification path. When either endpoint detects an error, it immediately alerts its peer via a separate bootstrap network, using MPI over a non-datapath NIC. This prevents peers from spinning for a long time on dead connections, reducing detection time from minutes to milliseconds.

\subsection{Precise Fault Localization}
A connection failure may stem from either endpoint, such as faulty NIC hardware, driver issues, or the physical link, but RDMA exposes only coarse transport errors (e.g., retry-exceeded), which lack the granularity to identify the impaired side. Localization therefore focuses on determining whether the failure is \textit{local}, \textit{remote}, or \textit{bilateral} to guide subsequent schedule optimization. 

\textbf{Our approach:}
R2CC introduces dedicated probe QP pools, isolated from data paths, to localize failures. When an error is detected, R2CC performs \textit{three-point triangulation} (for communicators spanning $\ge$3 servers): both endpoints and an auxiliary NIC on a third server of the same communicator issue zero-byte RDMA Writes as probes. A failed NIC produces immediate local probe errors, while its peer observes timeouts; a broken link yields timeouts at both endpoints, with the auxiliary NIC distinguishing single-endpoint vs. dual-endpoint impairment. By correlating probe outcomes, R2CC precisely identifies the fault and broadcasts it to all ranks via the OOB channel. For communicators spanning only two servers, R2CC directly marks both endpoints of the failed path unavailable and resumes on the backup connection. R2CC also periodically reprobes to detect component recovery (e.g., NIC resets, cable fixes), adapting probe frequency and strategy based on observed failure and recovery patterns.

\subsection{Live Migration for Failure Mitigation}
\begin{figure}
    \centering
    \includegraphics[width=1\linewidth]{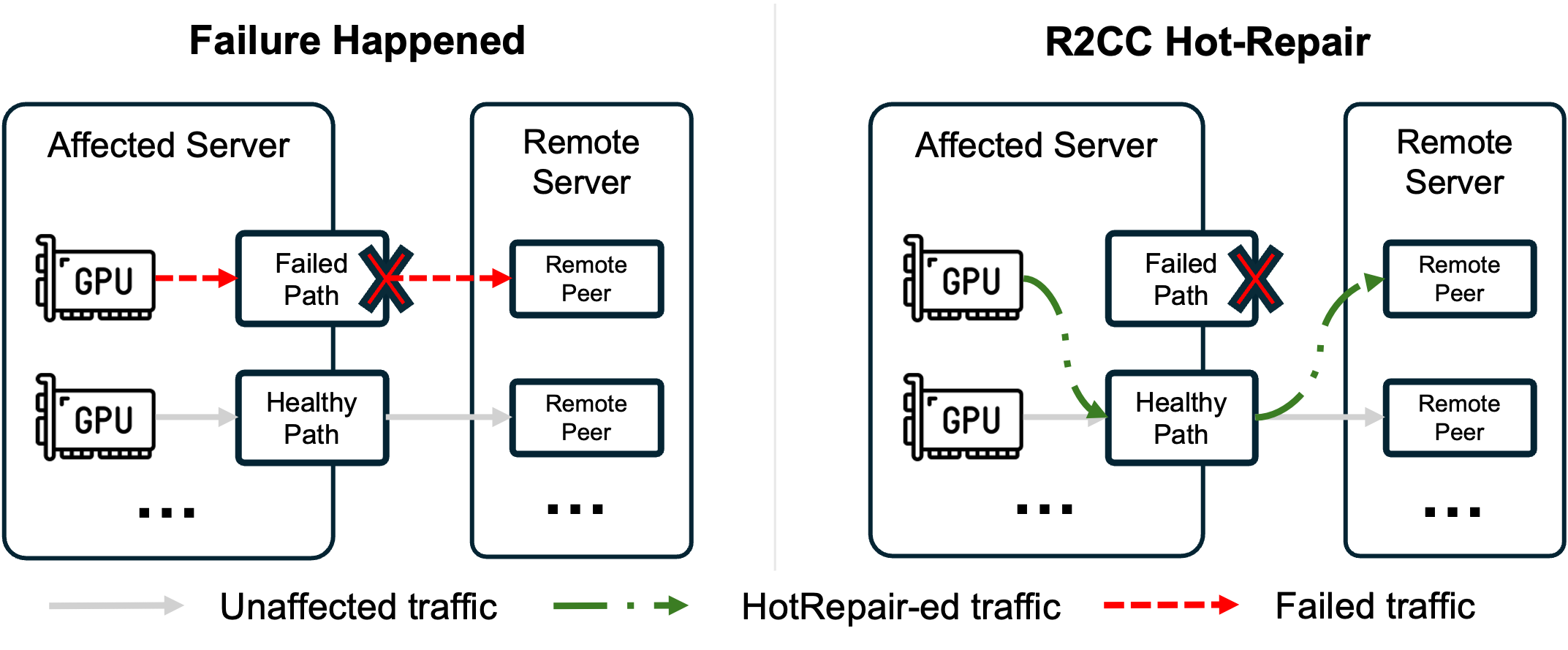}
    \vspace{-10pt}
    \tightcaption{\textbf{R2CC-HotRepair.} Affected traffic is switched to a pre-established healthy backup path after a failure.}
    \label{fig:new-hotrepair}
    \vspace{-2mm}
\end{figure}

After detecting and localizing a failure, the system must quickly resume collectives using the remaining healthy links (Figure~\ref{fig:new-hotrepair}). This requires fast, lossless connection migration so in-flight data is preserved with two challenges: (1) backup NICs cannot access GPU buffers unless those buffers were pre-registered with them, and (2) both sender and receiver must maintain data integrity during migration.

\paragraph{Technique I: Eager provisioning of the failover fast path.} R2CC eagerly provisions the failover fast path at communicator creation. This provisioning has two scopes: \emph{GPU buffer multi-registration} is a local memory-mapping decision, whereas \emph{backup QP pre-establishment} is a neighbor-edge decision.

\textit{GPU buffer multi-registration (local NIC scope).} An RDMA NIC can access a GPU buffer only if that buffer is registered with that NIC. R2CC therefore multi-registers each active communication buffer with the local NICs that may serve as immediate failover candidates. This decision is purely local and does not scale with the number of remote ranks or total cluster size. Extra registrations add only NIC-side translation metadata rather than duplicating buffer data. On our 8$\times$H100, 8-NIC testbed, pre-registering active communication buffers across all local NICs adds roughly 72\,ms to communicator initialization. Dynamic registration on the failure path is uncommon in practice because training frameworks typically reuse persistent communication buffers across iterations, and CUDA Graph replay further requires fixed memory addresses~\cite{cuda-graphs-memory-issues}.

\textit{Backup QPs (neighbor-edge scope).} R2CC pre-establishes backup QPs and peer connections only for existing neighbor edges, keeping their count on the same order as the transport state already required by the original topology rather than introducing a new cluster-wide scaling factor. Because backup QPs carry no steady-state traffic, they impose no steady-state data-plane overhead. On our 8$\times$H100, 8-NIC testbed, pre-establishing all backup QPs and peer connections adds 120\,ms and about 32\,MB of memory per GPU to communicator initialization. This bounded host/NIC-side cost is an explicit latency--memory tradeoff: it keeps connection setup off the in-flight repair path, reducing hot repair to selecting an already established backup connection and replaying the suffix. Rare neighbor edges introduced by post-failure optimization (Appendix~\ref{sec:multi-failure}) are provisioned outside this path.

\paragraph{Technique II: Protocol-aware suffix replay.} Pre-provisioning makes a backup path immediately available, but safe hot repair still requires the two endpoints to reconcile which suffix can be replayed without violating CCL protocol correctness. Upon detecting a failure, R2CC first freezes the recycle of CCL-managed GPU ring-buffer step slots. The GPU is not stopped: it may continue consuming any data already visible to it, but a slot under recovery cannot be handed to a later step until the replay boundary is reconciled. R2CC also keeps any sender-side staging slots needed for replay from being reused.

The two endpoints then exchange local progress over the OOB path and converge on a conservative replay boundary. Under the Simple protocol, where the GPU consumes a step only after the corresponding notification, the boundary is the receiver's last work completion. For LL and LL128, the boundary is the receiver's last fully consumed GPU step. The GPU may consume a step incrementally as LL flags become visible, but the step advances only after the entire step has been consumed. Steps at or below the boundary are not replayed, while steps after the boundary are replayed only while their slots remain frozen. Thus replay either completes an unfinished step in the same slot while that slot still belongs to the step, or skips a fully consumed step; it never writes an old step into a slot already recycled for a later step. Once the suffix catches up, R2CC releases the recycle gate and normal CCL progress resumes.

\subsection{Degradation Signaling and Restore}
\label{sec:signal-restore}

As R2CC keeps NCCL's collective API unchanged, network degradation is exposed through a separate query, \texttt{r2ccCommGetDegradationState()}, scoped to the communicator. The framework can poll this query at its own control points and chosen frequency, typically between iterations or requests. Each call returns the degradation epoch, the current state, the affected path, and the residual bandwidth, allowing the upper layer to decide whether to reschedule, throttle, or escalate to a heavier-weight recovery.

Restore is off the hot path. After failover, the current collective continues on the backup path while R2CC periodically reprobes the affected path. Once the original path becomes usable again, R2CC revalidates or re-establishes the primary path and switches subsequent collectives back at collective boundaries, so the primary and backup connections never access the buffer concurrently. It then reverts the degraded topology adaptation and updates the degradation signal exposed to the framework.

\section{Failure-aware Schedule Optimization}
\label{sec:single-failure}

When a failure is hot repaired, all in-flight messages are migrated from the failed NIC to a backup connection, which must now share bandwidth with other GPUs. If the collective continues to use its original schedule, all traffic formerly on the failed NIC now flows into the backup NIC, effectively doubling its load and significantly slowing communication.

We show that such a performance bottleneck can be \textit{largely} mitigated via a failure-aware design for all collectives: (1) \textit{R2CC-Balance} is a simple but generic load-balancing technique on top of NCCL's existing schedules. It evenly distributes traffic across the remaining healthy NICs and works for all collectives. (2) \textit{R2CC-AllReduce} is a novel AllReduce schedule that further improves throughput under failures.

\begin{figure}
    \centering
    \includegraphics[width=0.97\linewidth]{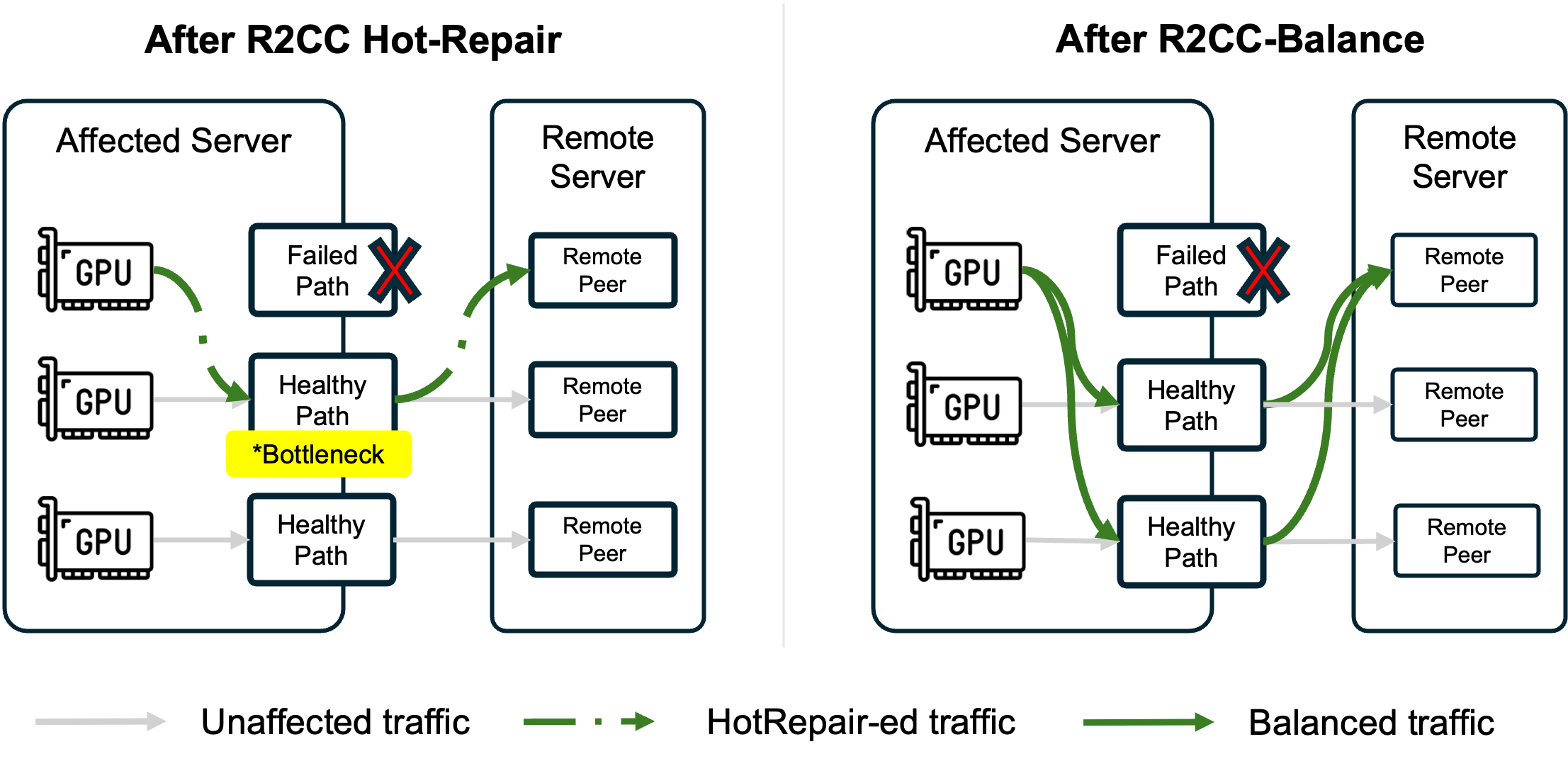}
    \tightcaption{\textbf{R2CC-Balance.} After HotRepair, R2CC redistributes the repaired traffic across the remaining healthy paths to avoid overloading a single path.}
    \label{fig:new-balance}
    \vspace{-3mm}
\end{figure}

\subsection{R2CC Load Balancing}

\paragraph{R2CC-Balance.} As shown in Figure~\ref{fig:new-balance}, our first technique is a NIC-level load-balancing strategy that leaves NCCL collective algorithms unchanged. R2CC-Balance treats NCCL schedules as fixed and intervenes only at the network layer, making it applicable to all collective types. For each collective invocation, NCCL's algorithm (ring, tree, or hierarchical variants using NVSwitch multicast) determines the participating GPUs and the amount of inter-server data each server $i$ must send, denoted $D_i$. Rather than redesigning the collective, R2CC-Balance exploits one remaining degree of freedom: when a NIC or link on server $i$ fails, it redistributes the portion of $D_i$ that would have used the failed NIC across the healthy NICs in proportion to their available bandwidth.

\paragraph{PXN- and NUMA-aware load balancing.}
When rerouting traffic from a GPU to a backup NIC, R2CC can either send it directly over PCIe or relay it via NVLink using GPU proxy forwarding (PXN). R2CC selects between these options on a per-flow basis using a topology-aware policy. A failed NIC frees its PCIe lane from the affinity GPU, creating additional headroom that R2CC can exploit for detoured flows. Thus, R2CC prioritizes {\bf direct PCIe forwarding}: if the backup NIC is on the same NUMA node and its PCIe path has sufficient bandwidth headroom, the source GPU forwards traffic via PCIe. If the target NIC lies on a different NUMA node, R2CC compares the cost of traversing the CPU interconnect (e.g., QPI/UPI) with the NVLink headroom available for PXN. It then chooses the lower-cost path: either PCIe plus CPU-interconnect forwarding, or {\bf PXN forwarding} to a proxy GPU co-located with the target NIC.

\paragraph{Overhead Analysis.} NCCL's existing schedules already drive each server's cross-server traffic close to the semantic minimum for core collectives. For total data size $D_{\text{total}}$: a ReduceScatter retains only a $1/n$ shard and must send $\frac{n-1}{n}D_{\text{total}}$; an AllGather must receive the same amount; and in a Broadcast, the root sends $D_{\text{total}}$ while all others receive it. NCCL's ring algorithm realizes these lower bounds in homogeneous systems.

After a NIC failure, these volumes do not change: server $i$ must still exchange $D_i$ bytes, even though its aggregate NIC bandwidth is reduced to $B_i^{\text{rem}}$. Because these collectives inherently require $\Theta(D_{\text{total}})$ cross-server traffic, no schedule can substantially reduce $D_i$; the only leverage is how efficiently the remaining bandwidth is used. R2CC-Balance therefore redistributes the portion of $D_i$ originally routed through the failed NIC across the healthy NICs, allowing their combined throughput to approach $B_i^{\text{rem}}$. So completion time is dictated primarily by the reduced capacity of the slowest server.

\subsection{R2CC AllReduce}

Ring-based AllReduce is typically realized as a two-stage protocol (e.g., in NCCL and RCCL): A ReduceScatter over all GPUs followed by an AllGather over all GPUs, while tree-based AllReduce corresponds to a Reduction on a tree followed by a Broadcast on the same tree.

\begin{figure}[t]
   \centering
   \includegraphics[width=1\linewidth]{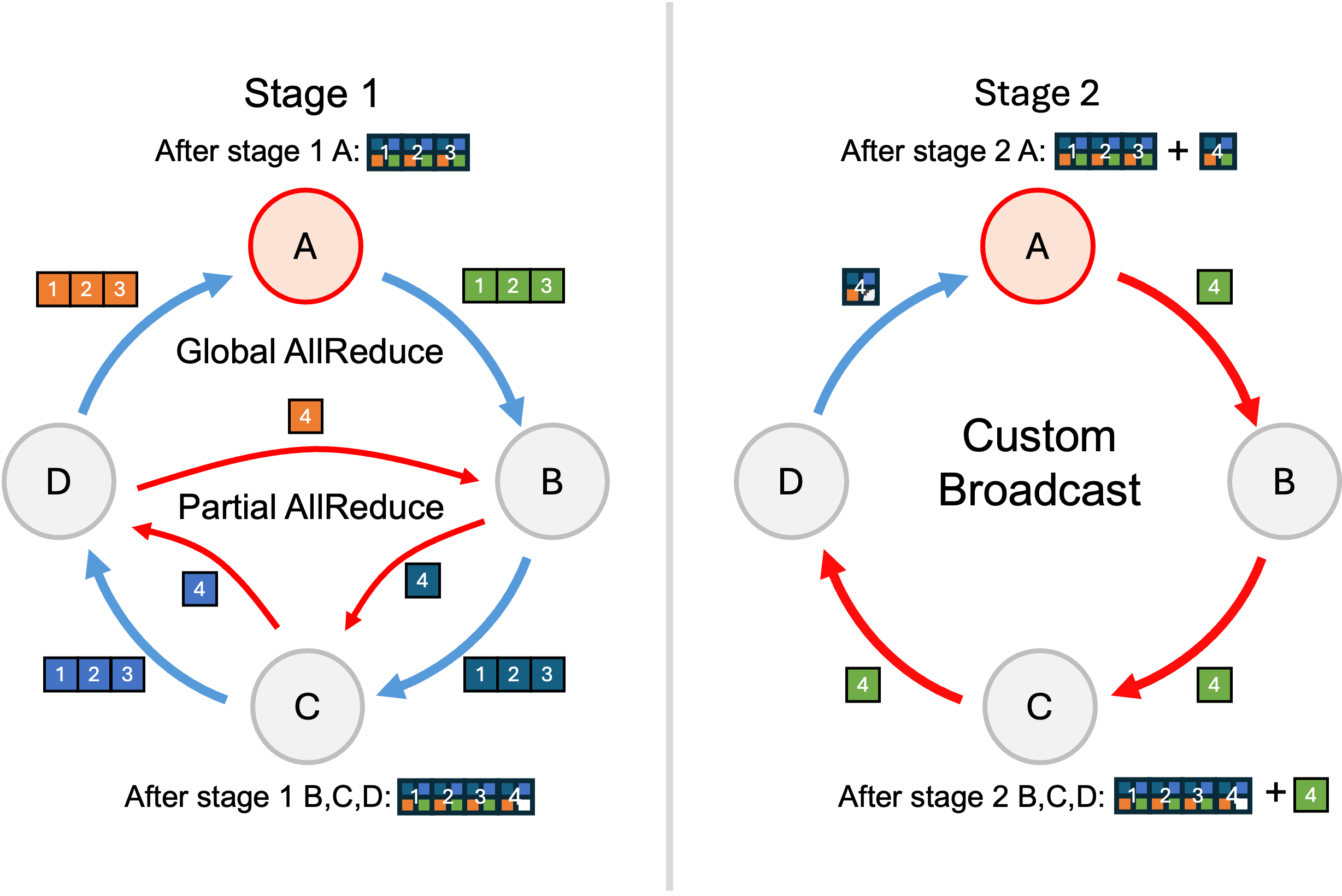}
   \vspace{-15pt}
    \tightcaption{\textbf{R2CC-AllReduce reduces the degraded node’s load from $2D$ to $7/4D$.} In this 4-node example with lost-bandwidth fraction $1/4$, Stage~1 runs a global AllReduce with all 4 nodes on $3/4D$ and a partial AllReduce with 3 healthy nodes on $1/4D$, and Stage~2 broadcasts the $1/4D$ partial result back. Thus node $A$ carries $2\!\cdot\!3/4D + 1/4D = 7/4D$. Since the data and bandwidth in Stage 1 are proportional, the total completion times are nearly the same for each rank. }
    \label{fig:algorithm}
\end{figure}

To AllReduce $D$ size data per GPU, in ReduceScatter operation, a server must transmit all $D$ size data via its NIC, excluding the portion reduced onto itself; conversely, in AllGather, it must receive at least $D$ size data from the others. Because the two operations execute sequentially rather than concurrently, the peak load on a server consists of transmitting roughly $D$ during ReduceScatter and receiving roughly $D$ during AllGather, leading to a combined data volume of approximately $2D$.

In contrast, for an AllReduce operation, a key insight is that {\bf a server can wait for other nodes to complete a \textit{partial AllReduce} that excludes itself, and subsequently send and receive only the data it requires (broadcast) to satisfy the full operation demand}. Moreover, partial AllReduce can proportionally reduce the load on a specific server such as a server with a failed NIC, to improve the overall AllReduce performance. In practice, the failure server node retains some healthy NICs; we can allow it to participate in the global AllReduce across all ranks over a
subset of the data, while the remaining data are handled by a
partial AllReduce that excludes the failure node.

\paragraph{R2CC-AllReduce.} Based on the above insights, we propose a novel AllReduce algorithm (Figure~\ref{fig:algorithm}). When NIC failures induce inter-node bandwidth heterogeneity, we partition the AllReduce into two stages:

(1) \textit{The first stage} executes a global AllReduce and a partial AllReduce concurrently. The global AllReduce spans all servers, including the one with the failed NIC, and is therefore throttled by the reduced effective bandwidth of the affected server. In parallel, the partial AllReduce excludes the failure node and utilizes the remaining capacity of the healthy servers to process its share of data at full speed.

(2) \textit{The second stage} executes a tailored broadcast that completes the partial-AllReduce plus broadcast path. This process comprises a broadcast initiated from the failure server node, a pipelined ring broadcast across the healthy servers, and the final delivery of the partial-AllReduce result from the last node in the ring back to the failure node, ensuring that all servers obtain the full correct AllReduce result.

\smallskip\noindent\textbf{Data partition and implementation.} Let $D$ be the per-node AllReduce data size, $B$ the per-node bandwidth, $n \geq 2$ the number of nodes, $X \in (0, 1)$ the fraction of bandwidth lost on the failed node, and $Y \in [0, 1]$ the fraction of data routed through the partial AllReduce. In Stage~1, a global AllReduce over all $n$ nodes on $(1-Y)D$ bottlenecked by the failed node's $(1-X)B$ runs concurrently with a partial AllReduce over the $n-1$ healthy nodes on $YD$ using their reserved $XB$; Stage~2 broadcasts the $YD$ partition back to the failed node over its $(1-X)B$:
\begin{align*}
T_1(Y) &= \frac{2(n-1)}{n} \cdot \frac{(1-Y)D}{(1-X)B}, \\
T_2(Y) &= \frac{2(n-2)}{n-1} \cdot \frac{YD}{XB}, \\
T_3(Y) &= \frac{YD}{(1-X)B}.
\end{align*}
The total completion time $T(Y) = \max(T_1(Y), T_2(Y)) + T_3(Y)$ is minimized at
$$Y^* = \frac{(n-1)^2 X}{n^2 - 2n + X}.$$
Because $Y^*$ produces awkward byte offsets that complicate buffer management and memory alignment, we instead adopt the simpler rule $Y = X$, whose approximation gap satisfies
$$\frac{T(X) - T(Y^*)}{T(Y^*)} < Y^* - X \leq \frac{1}{4n(n-2)},$$
which decays as $O(1/n^2)$ and reaches at most $1.2\%$ at $n = 4$. The full derivation and per-$(n, X)$ values are given in Appendix~\ref{appendix:allreduce-math}. R2CC-AllReduce therefore adopts $Y = X$, and we use $T(X)$ to denote its completion time hereafter.

\begin{figure}[t]
    \centering
    \includegraphics[width=0.85\columnwidth]{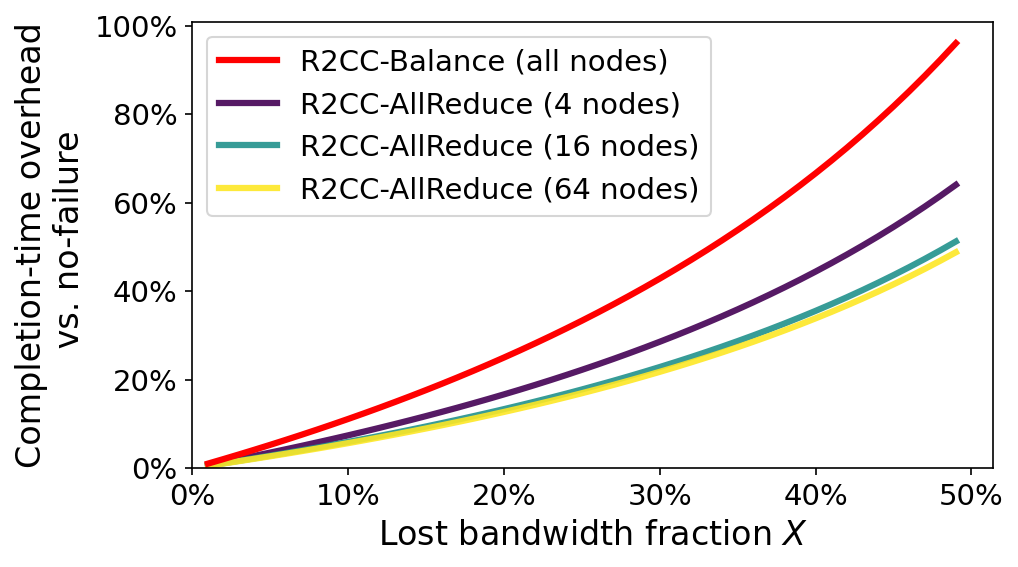}
    \vspace{-5pt}
    \tightcaption{\textbf{Overhead Analysis.} Completion-time overhead vs.\ no-failure as a function of lost bandwidth fraction $X$.}
    \label{fig:overhead-compare}
    \vspace{-2mm}
\end{figure}

\textbf{Comparison with no-failure time.} To quantify the failure-induced overhead of each variant, we compare against the no-failure ring AllReduce time $T_{\text{nf}} = \frac{2(n-1)D}{nB}$. Direct computation gives
$$\frac{T_{\text{Balance}} - T_{\text{nf}}}{T_{\text{nf}}} = \frac{X}{1-X}, \qquad \frac{T(X) - T_{\text{nf}}}{T_{\text{nf}}} = \frac{nX}{2(n-1)(1-X)}.$$
The ratio of Balance's overhead to R2CC-AllReduce's overhead is exactly $\frac{2(n-1)}{n}$, the ring AllReduce efficiency factor, which approaches $2$ as $n \to \infty$. Thus R2CC-AllReduce asymptotically halves the failure-induced overhead relative to R2CC-Balance. Figure~\ref{fig:overhead-compare} plots both overheads across the full range of $X$: at $X = 50\%$ and $n = 16$, R2CC-Balance incurs a 100\% overhead while R2CC-AllReduce with $Y = X$ cuts it to 53.3\%.

\paragraph{Optimize Scheduling Under Multi-Failure.} Beyond the single-failure scenario, handling multi-failure case is discussed in Appendix~\ref{sec:multi-failure}.

\section{Implementation}
We implement R2CC as a drop-in extension of 3K lines of C++ code to NCCL 2.23.4, allowing existing ML frameworks to leverage it without any modifications. Our implementation integrates R2CC strategies into the host transport layer and collective scheduling logic, and builds R2CC-AllReduce from standard NCCL collectives on sub-communicators.

\smallskip\noindent\textbf{Transport hooks and backup paths.}
R2CC intercepts NCCL's communicator initialization at the \texttt{ncclNet} plugin layer to establish backup connections alongside the primary connection. We extend the memory registration path to register each GPU buffer with all NICs of the server, and augment the connection setup logic to create additional queue pairs over backup NICs. The resulting handles are organized into a per-channel failover list ordered by PCIe distance to the source GPU, allowing the failure recovery logic to select the topologically closest healthy NIC.

\smallskip\noindent\textbf{Failure detection and live migration.}
R2CC augments the net plugin and proxy send receive progress loops to handle network failures without crashing the application process. The modified InfiniBand plugin monitors Completion Queue (CQ) and Queue Pair (QP) errors, reporting them as failed requests to the proxy. Upon error, the plugin issues a lightweight RDMA probe and exchanges messages via the bootstrap network to confirm the failure location. Once a failure is confirmed, the proxy rolls the channel back to the last chunk with a confirmed completion, purges outstanding work requests, transitions the channel to the backup connection, and reissues the remaining chunks.

\smallskip\noindent\textbf{R2CC-Balance and R2CC-AllReduce.}
R2CC-Balance is integrated into NCCL's enqueue and planning logic. Before NCCL selects an algorithm and instantiates rings or trees, R2CC inspects the health status records of NICs on each node. The planner then redistributes the affected channels across the remaining healthy NICs on the same node. R2CC-AllReduce creates two sub-communicators with \texttt{ncclCommSplit} after repair. Most of the data runs a global AllReduce across all nodes, while the rest runs a Partial AllReduce on the sub-communicator that excludes the degraded node. The degraded node's share of that part is then reduced onto a healthy helper rank and broadcast back in pipelined chunks. These collectives run on separate CUDA streams ordered by events, so the phases overlap without custom kernels. HotRepair, R2CC-Balance, and R2CC-AllReduce all support CUDA Graphs (Appendix~\ref{app:discussion}).

\section{Evaluation}

We evaluate R2CC extensively on a real testbed and large-scale simulations. We find that:
\begin{packeditemize}
    \item R2CC achieves less than 1.1\% overhead on training and less than 3\% on inference when failure happens, reducing overhead by up to 92\% compared to training baselines and up to 98\% compared to inference baselines (§\ref{exp:training}, §\ref{exp:inference}).
    \item R2CC demonstrates robust scalability, incurring only 4.3\% overhead with 10 concurrent failures across a simulated 512-GPU cluster (§\ref{exp:training}).
    \item R2CC maintains up to 93\% of baseline throughput for collective operations during failures (§\ref{exp:overhead}).
\end{packeditemize}

\subsection{Methodology}
\label{sec:setup}
\textbf{Testbed.} Our physical testbed consists of four server nodes interconnected via InfiniBand fabric. Each node is equipped with 8 NVIDIA H100-80GB SXM5 GPUs, and 8 ConnectX-7 400Gbps InfiniBand NICs\footnote{InfiniBand SHARP is not enabled on our shared testbed; we discuss R2CC with SHARP in Appendix~\ref{app:discussion}.}. The intra-node GPUs support NVLink 4.0, providing 900GB/s bidirectional bandwidth. We use all four nodes for data-parallel training and the microbenchmarks, and two nodes for the TP=8, PP=2 training and inference runs. This configuration covers typical distributed training paradigms, allowing us to evaluate R2CC under realistic network conditions and failure scenarios.

For simulation experiments, we use SimAI~\cite{wang2025simai}, a high-fidelity ML training simulator that provides end-to-end training performance estimation, packet-level network modeling, and precise collective communication timing. SimAI has been validated against real-system performance and adopted in recent studies~\cite{qin2025phantora,gui2025accelerating,huang2025netzip}. We configure SimAI with NVIDIA's Spectrum-X RoCE-v2 rail-optimized topology, with each simulated server equipped with eight NVIDIA A100 GPUs and eight 200\,Gbps NICs. To evaluate the scalability of R2CC, we simulate clusters of 32 to 1024 GPUs, beyond the scale of our physical testbed.

\smallskip\noindent\textbf{Baselines.} For training experiments, we compile PyTorch v2.10.0 with NCCL v2.23.4 and with R2CC, and compare with these approaches: (1) {\bf Vanilla NCCL}~\cite{NCCL}, which crashes on NIC failure; (2) {\bf AdapCC}~\cite{zhao2024adapcc}, which excludes failed GPUs; (3) {\bf SHIFT}~\cite{shift}, which tolerates NIC failures by transparently failing over RDMA traffic to another intra-host NIC. (4) {\bf ICCL}~\cite{chen2025efficient}, which tolerates NIC port failures by always switching communication to a backup queue pair.

For inference experiments, we compile NCCL v2.23.4 and R2CC and build vLLM v1 on top. Because NCCL lacks fault tolerance, we implement two standard mitigation strategies: (1) {\bf Restart Server}, which incurs a measured 35-second restart delay, and (2) {\bf Reroute Request}, which redirects affected queries to an alternative server that must absorb the doubled load. We also evaluate {\bf DéjàVu}~\cite{ICML2024}, a fault-tolerant inference framework with KV-cache replication, compiled with both vanilla NCCL and R2CC to isolate R2CC's contribution.

For microbenchmarks, we use NCCL-tests v2.17.1 on all four nodes to compare vanilla NCCL and R2CC for collective and point-to-point operations under normal and single-NIC failure conditions.

\subsection{Training Resilience}
\label{exp:training}

\begin{figure}[t]
  \centering

  \includegraphics[width=\columnwidth]{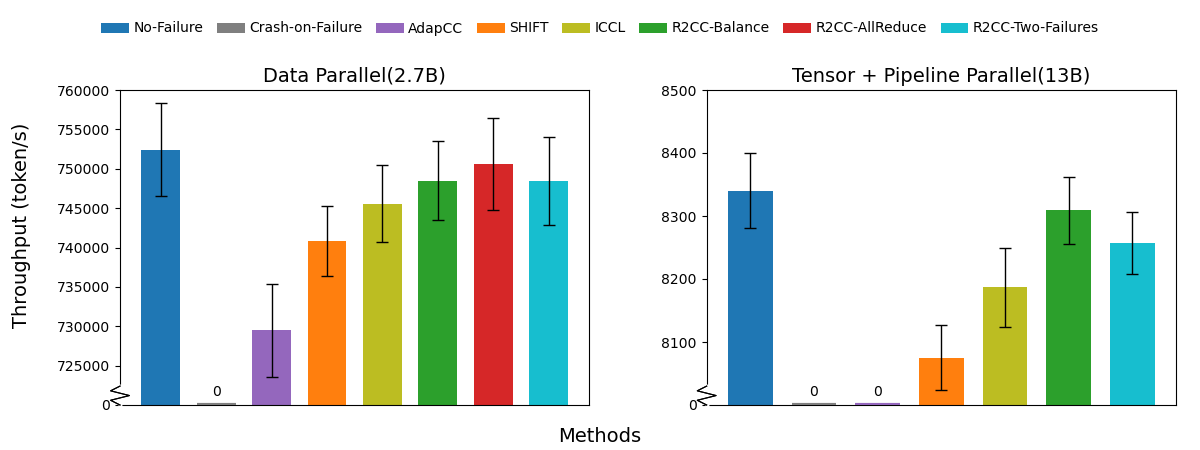}
  \vspace{-5mm}
  \tightcaption{Megatron Training Performance Evaluation}
  \label{fig:megatron}
  \vspace{-3mm}
\end{figure}

In this section, we demonstrate R2CC's effectiveness in maintaining training continuity when failures occur. We use Megatron-LM\cite{narayanan2021efficient} as our training benchmark to compare against baselines.

\smallskip\noindent\textbf{Physical testbed evaluation.}
We measure token-per-second training throughput (Figure \ref{fig:megatron}) and overhead relative to a no-failure baseline for GPT-3 2.7B and 13B models under two multi-node configurations:
(1) DP=32 on the 2.7B model, where inter-node all-reduce dominates;
(2) TP=8, PP=2 on the 13B model, where pipeline stages span nodes and induce point-to-point activations.

Under DP=32, vanilla NCCL fails completely. Among fault-tolerant methods, R2CC-AllReduce achieves 750,627 tokens/s, just 0.24\% overhead vs.\ the no-failure throughput of 752,405 tokens/s. R2CC-Balance incurs 0.43\%, SHIFT and ICCL incur 1.57\% and 0.92\% respectively. AdapCC is the slowest (3.14\%) due to reduced compute capacity.

For TP=8, PP=2 on the 13B model, R2CC-Balance performs best, with only 0.38\% overhead (8,309 vs. 8,340 tokens/s). SHIFT and ICCL incur 3.28\% and 1.88\% respectively. AdapCC cannot operate in this setting (0 tokens/s) because removing a rank violates tensor/pipeline partitioning constraints.

AdapCC can exclude failed GPUs only if the failure occurs before a collective begins, so mid-operation faults still crash the job and excluded GPUs cause gradient loss. R2CC instead reroutes traffic through healthy NICs mid-operation and preserves full-rank, lossless execution.

To assess resilience under more severe conditions, we inject two simultaneous NIC failures on one node. In DP=32, R2CC-Two-Failures sustains 749,134 tokens/s (0.44\% overhead). In TP=8, PP=2, it delivers 8,257 tokens/s (1.01\% overhead). In both cases, training proceeds with near-baseline efficiency, demonstrating R2CC's robustness to multiple failures.

\smallskip\noindent\textbf{Training quality.} On a separate three-server testbed with six NVIDIA Tesla V100S GPUs, we train GPT-2~\cite{radford2019language} (124M) through a NIC-port failure and compare it with NCCL runs. Failure-free R2CC matches NCCL bit for bit, and the collective repaired by HotRepair yields a bit-identical gradient. The degraded schedules change only the order in which gradients are reduced, and since floating-point addition is not associative, this introduces rounding-level deviations: the per-update training loss differs by at most $2.4\times10^{-3}$ nats/token and final test perplexity by at most 0.023\% (Appendix~\ref{app:quality}).

\begin{figure}[t]
    \centering

    \begin{subfigure}[b]{0.48\columnwidth}
        \includegraphics[width=\textwidth]{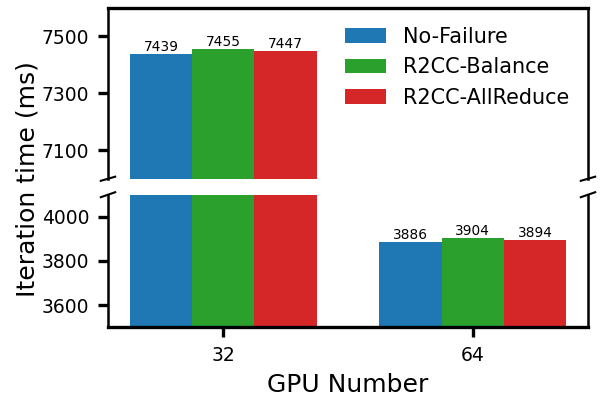}
        \caption{Small Scale Training.}
        \label{fig:training-small}
    \end{subfigure}
    \hfill
    \begin{subfigure}[b]{0.48\columnwidth}
        \includegraphics[width=\textwidth]{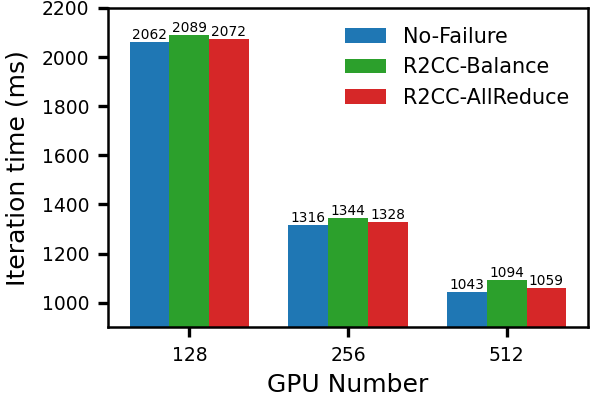}
        \caption{Large Scale Training.}
        \label{fig:training-large}
    \end{subfigure}

    \vskip\baselineskip

    \begin{subfigure}[b]{0.48\columnwidth}
        \includegraphics[width=\textwidth]{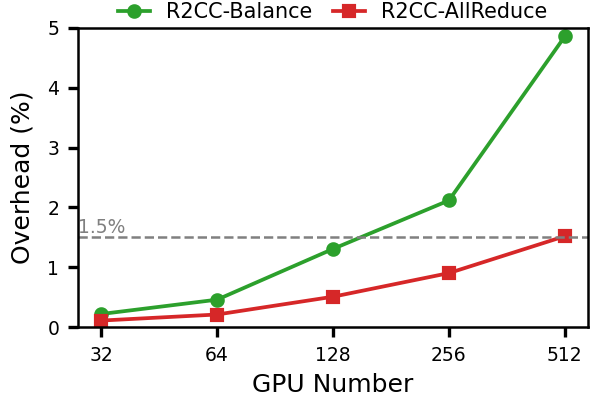}
        \caption{Strategies Overhead.}
        \label{fig:training-e2eoverhead}
    \end{subfigure}
    \hfill
    \begin{subfigure}[b]{0.48\columnwidth}
        \includegraphics[width=\textwidth]{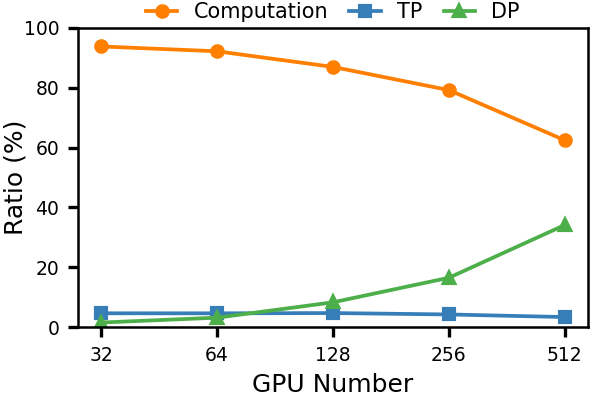}
        \caption{Time Composition.}
        \label{fig:training-e2eratio}
    \end{subfigure}

    \tightcaption{Simulated 7B model training on 4--64 8$\times$A100 servers.}
    \label{fig:training-all}
\end{figure}

\begin{figure}[t]
    \centering
    \begin{minipage}[t]{0.49\linewidth}
        \centering
        \includegraphics[width=\linewidth]{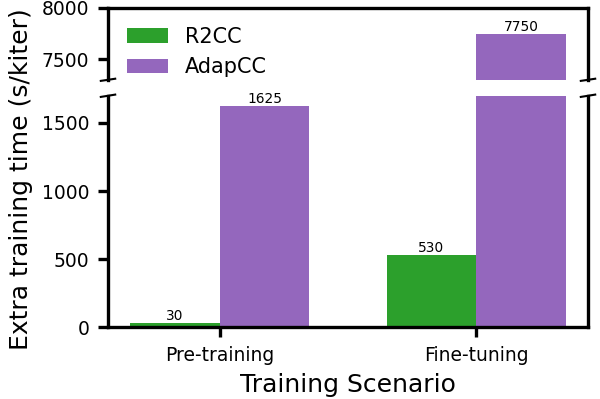}
        \vspace{-10pt}
        \tightcaption{Extra training time in R2CC vs. AdapCC.}
        \label{fig:extra_training_time}
    \end{minipage}
    \hfill
    \begin{minipage}[t]{0.49\linewidth}
        \centering
        \includegraphics[width=\linewidth]{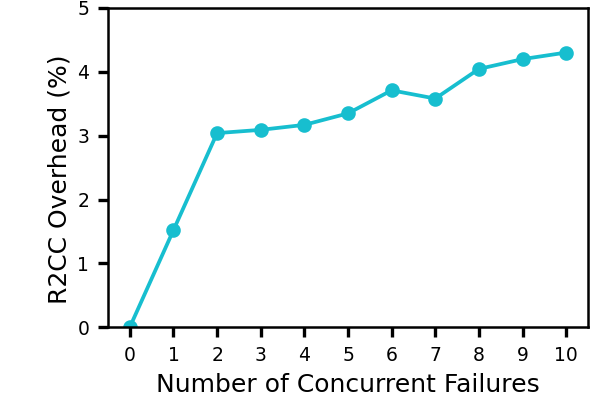}
        \vspace{-10pt}
        \tightcaption{Multi-failure training overhead.}
        \label{fig:training_multi_failures}
    \end{minipage}
\end{figure}

\smallskip\noindent\textbf{Large scale training simulation under failures.}
For a single failure, we use SimAI to simulate a 7B-parameter model with global batch size 512, from 4 to 64 servers each with 8 GPUs. Within each server, GPUs are partitioned into two TP groups. Figure~\ref{fig:training-e2eoverhead} shows that R2CC-AllReduce consistently maintains an overhead of less than 1.5\% despite a 12.5\% bandwidth loss, significantly outperforming the R2CC-Balance strategy, which sees overhead rise to 5\% at larger scales. Overheads increase as the number of GPUs increases, because communication ratio increases when per GPU computation batch decreases, as shown in Figure~\ref{fig:training-e2eratio}.
Figure~\ref{fig:extra_training_time} shows pre-training a 175B-parameter model with 1,024 GPUs (TP=8, PP=8, DP=16) and RLHF fine-tuning with DeepSpeed-Chat's RLHF~\cite{yao2023deepspeedchateasyfastaffordable} on 64 GPUs (TP=8, PP=1, DP=8) with Fully Sharded Data Parallelism (FSDP). In this production scenario, R2CC proved vastly superior to the AdapCC framework reducing failure-induced training time by approximately 54$\times$ and 15$\times$, respectively.

We also extend our evaluation to scenarios with multiple concurrent failures. Unlike single-failure cases where the impact is deterministic, the performance degradation in multi-failure scenarios depends heavily on the spatial distribution of faults. We randomly inject $k$ failures across 64 servers (512 GPUs), varying $k$ from 1 to 10. We conduct a Monte Carlo simulation with 50 random failure patterns for each $k$. Figure~\ref{fig:training_multi_failures} shows the average iteration time overhead for the 7B model training as the number of failures increases. R2CC demonstrates exceptional resilience: the overhead grows sub-linearly from 1.5\% with a single failure to only 4.3\% with 10 concurrent failures. When multiple failures concentrate on a single server, that server becomes a bandwidth bottleneck; when failures scatter across the cluster, the impact is amortized and no single server dominates the critical path. This confirms that R2CC's load-balancing and algorithmic optimizations effectively bound performance degradation even under severe failure scenarios.

\subsection{Inference Resilience}\label{exp:inference}

\begin{figure}
    \centering
    \includegraphics[width=1\linewidth]{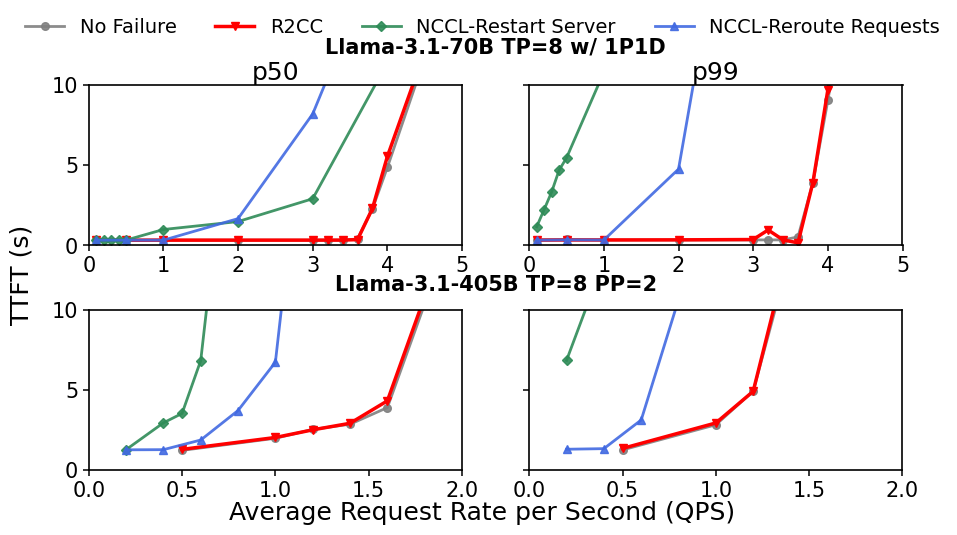}
    \vspace{-10pt}
    \tightcaption{p50/p99 TTFT vs QPS under NIC failure for vLLM}
    \label{fig:vllm_results}
    \vspace{-3mm}
\end{figure}

\begin{figure}[t]
    \centering
    \includegraphics[width=1\linewidth]{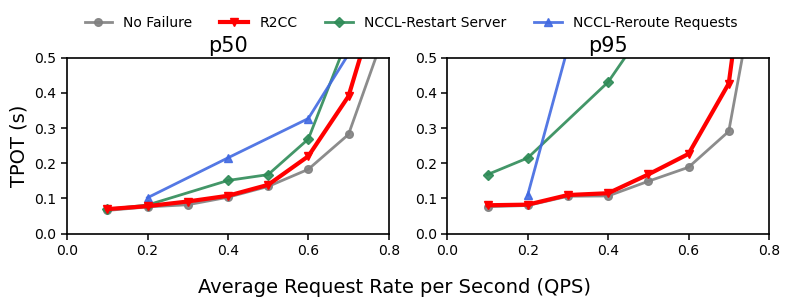}
    \vspace{-10pt}
    \tightcaption{
    p50/p95 TPOT under NIC failure for serving Llama-3.1-405B with TP = 8, PP = 2.}
    \label{fig:tpot}
    \vspace{-3mm}
\end{figure}

In this section, we demonstrate how R2CC leverages network redundancy to significantly benefit end-to-end inference performance
when network failures occur. We evaluate R2CC using vLLM\cite{kwon2023efficient}, a widely adopted production inference framework, and compare it against two strategies relied upon by systems lacking transport-layer resilience: restarting the failed service and rerouting the requests to another server. Additionally, we integrate R2CC into DéjàVu\cite{ICML2024}, an existing fault-tolerant inference framework, to compare R2CC with its application-side fault tolerance.

\paragraph{vLLM Inference Performance.}
We use the vLLM v1 engine and focus on two multi-node configurations that exercise inter-node communication:
(i) tensor-parallelism with pipeline parallelism (TP=8, PP=2), and
(ii) TP=8 with prefill–decode (PD) disaggregation setting, which separates the compute-intensive prefill phase from the memory-intensive decode phase across different nodes.
These are vLLM configurations that require inter-node communication rather than being confined to a single node.

For each setup, we generate requests at fixed arrival rates. We measure different metrics tailored to each configuration's characteristics. For PD disaggregation, inter-node network failures primarily impact the prefill phase, making time-to-first-token (TTFT) our key metric since decoding proceeds without inter-node communication after the prefill node finishes transferring the KV cache. The pipeline-parallel configuration requires inter-node communication for every token generation, so we track both TTFT and time-per-output-token (TPOT) to capture the full performance impact.

Our experiments simulate inference patterns using 2000-token synthetic prompts, allowing models to generate outputs until reaching EOS or 256 tokens. Each 100-second experiment injects a single NIC failure at the midpoint (t=50s) to observe both transient and steady-state effects.

\paragraph{End-to-end latency (TTFT).}
Fig.~\ref{fig:vllm_results} plots TTFT as a function of the offered load for Llama-3.1-70B and 3.1-405B under different failure-handling strategies.
R2CC almost overlaps with the no-failure curve across models and percentiles (p50/p99),
indicating near-zero overhead under a NIC failure: before saturation, the TTFT overhead is within $\approx$0--0.6\% for the 70B model and 0.3--3\% for the 405B model.
Under a TTFT SLO of 5\,s, R2CC sustains up to 1.2--8.7$\times$ higher throughput than service restart and 1.6--1.9$\times$ higher throughput than request rerouting, while retaining $\approx$99–100\% of the no-failure capacity.

\paragraph{Streaming latency (TPOT).}
Fig.~\ref{fig:tpot} reports time per output token (TPOT) for the Llama-3.1-405B TP+PP configuration.
R2CC again closely tracks the no-failure baseline: both p50 and p95 TPOT increase smoothly with QPS and remain low until the system saturates, showing that R2CC preserves the streaming rate of generated tokens.
Before saturation, the TPOT overhead is within 3\%.
Under a p95 TPOT SLO of 0.4 s, R2CC sustains $\approx$1.9× and $\approx$2.6× higher throughput than the restart and rerouting strategies, respectively, while remaining within about 5\% of the no-failure capacity.

Appendix~\ref{multi-failure-inference} presents results under multiple failures.

\paragraph{Performance Comparison vs. DéjàVu.}
We follow DéjàVu's evaluation methodology and keep the application stack unchanged, varying only the communication layer. Using two workers with TP8 and PP2 parallelism, we evaluate the same models—OPT-66B and BLOOM-176B—under homogeneous requests (500-token prompt, 1500-token generation) and inject failure at decode step 800. As in their study, we report single-request cumulative latency. For comparison, we compile DéjàVu in two forms: (i) its original NCCL-based build and (ii) a version where NCCL is replaced by R2CC. This setup also demonstrates that R2CC integrates seamlessly without requiring any application-level modifications.

\begin{figure}[t]
    \centering
    \includegraphics[width=1\linewidth]{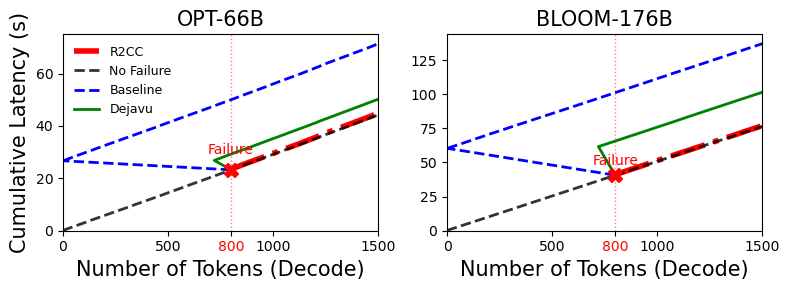}
    \vspace{-10pt}
    \tightcaption{Inference performance under failure compared to DéjàVu and non-fault-tolerant baseline.}
    \label{fig:dejavu}
\end{figure}

For DéjàVu and the non-fault-tolerant baseline, recovery is dominated by worker restart and reconnection delays. As shown in Figure~\ref{fig:dejavu}, the non-fault-tolerant baseline experiences severe latency inflation: $1.62\times$ for OPT-66B and 1.79$\times$ for BLOOM-176B—due to full request reprocessing. DéjàVu reduces this via neighbor KV-cache replication but still incurs a 14–33\% penalty (1.14$\times$–1.33$\times$) from bandwidth-heavy state reconstruction and added memory overhead. In contrast, R2CC's transparent connection migration avoids both restart and reconstruction, sustaining service with only 0.71–1.58\% overhead.
R2CC reduces recovery overhead by about 89--95\% for OPT-66B and 95--98\% for BLOOM-176B. Relative to the non-fault-tolerant baseline, the reduction is about 97--99\% for both models.

\subsection{Microbenchmarks}\label{exp:overhead}
\label{sec:microbenchmark}

We evaluate R2CC's primitive communication performance using NCCL-tests on four nodes of our physical testbed, measuring communication operations bus bandwidth across message sizes from 8B to 16GB under a failure scenario.

\begin{figure}[t]
    \centering
    \begin{minipage}[t]{0.48\linewidth}
        \centering
        \includegraphics[width=1\textwidth]{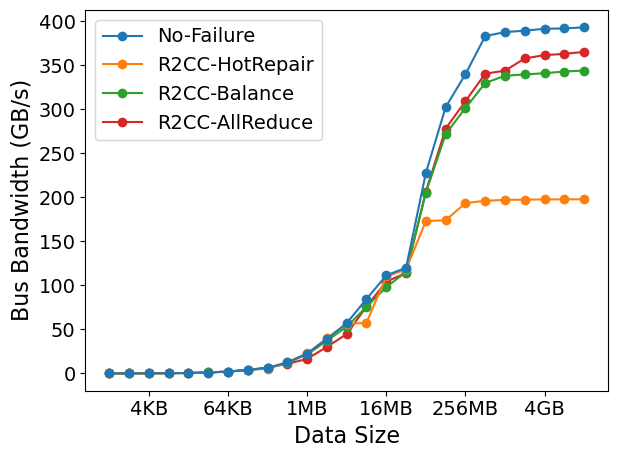}
        \vspace{-10pt}
        \tightcaption{AllReduce bandwidth vs. different message sizes using different R2CC configurations.}
        \label{fig:allreduce-h100}
    \end{minipage}
    \hfill
    \begin{minipage}[t]{0.48\linewidth}
        \centering
        \includegraphics[width=\linewidth]{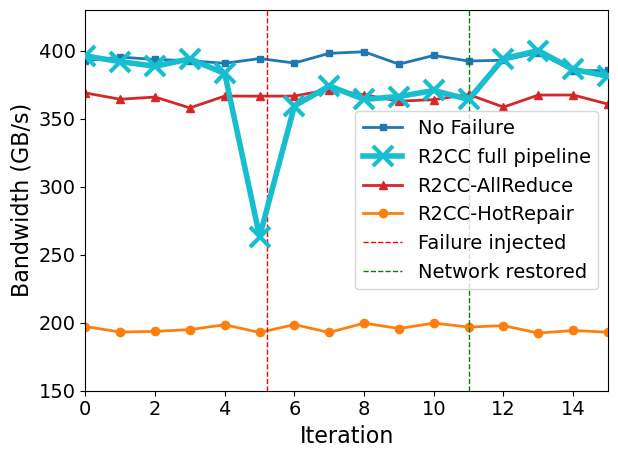}
        \vspace{-10pt}
        \tightcaption{AllReduce bandwidth at message size 1GB with R2CC pipeline for failover and restore.}
        \label{fig:tiaolaitiaoqu}
    \end{minipage}
\end{figure}

\paragraph{AllReduce Performance.}
Figure~\ref{fig:allreduce-h100} compares four configurations: vanilla NCCL with no failure, R2CC-HotRepair, R2CC-Balance, and R2CC-AllReduce. We also run the full R2CC pipeline with failures injected and fixed to measure R2CC's behavior with failover and restoration.

Without failure, vanilla NCCL achieves up to 393 GB/s for large messages. When a NIC fails (12.5\% total bandwidth loss), the R2CC-HotRepair approach suffers approximately 50\% throughput loss for large messages, as expected from the single-NIC bottleneck on the affected node.

R2CC's two strategies exhibit complementary strengths. For small messages ($<$32MB), R2CC-Balance achieves significantly lower latency, maintaining 88--96\% of normal throughput while R2CC-AllReduce ranges 73--94\% due to its data dependency coordination overhead. For medium messages (32MB--256MB), both strategies perform similarly at 88--96\% of normal throughput. For large messages ($\geq$512MB), R2CC-AllReduce achieves superior performance, around 88--93\% of normal throughput compared to Balance's 86--88\%, by reducing the data volume that the bandwidth-constrained node must transmit. These large-message ratios match the closed-form predictions from Section~\ref{sec:single-failure}: at $n = 4$ and $X = 1/8$, Balance throughput is expected to be $(1-X) = 87.5\%$ of baseline and R2CC-AllReduce throughput $\frac{2(n-1)(1-X)}{2(n-1)(1-X)+nX} = 91.3\%$, both falling within the measured ranges.

In practice, R2CC leverages an $\alpha$-$\beta$ performance model, using machine-specific link latency ($\alpha$) and bandwidth ($\beta$) parameters provided by NCCL along with the operation's data size, to dynamically select the optimal strategy at runtime. This approach ensures that the crossover point adapts to different hardware configurations and message sizes, rather than relying on a fixed threshold.

We evaluate the full R2CC pipeline from HotRepair to restore with a 1~GB message size nccl-tests AllReduce benchmark and compare its performance across different policies. As illustrated in Figure~\ref{fig:tiaolaitiaoqu}, we inject a network failure midway through iteration 5 and restore the connection at iteration 11. R2CC immediately repairs the in-flight collectives when the failure occurs, which causes a temporary dip in the average bandwidth for iteration 5. Starting from iteration 6, R2CC adapts to the optimal R2CC-AllReduce strategy, continuing the operation with only a 6.9\% overhead compared to the failure-free baseline. Once the network is restored at iteration 11, R2CC automatically reverts to the original AllReduce schedule. We also evaluate a case where a NIC fails mid-training between two scheduled checkpoints in Appendix~\ref{appendix:checkpoint}.

\paragraph{More results on other collectives.} Beyond AllReduce, we evaluate R2CC-Balance on AllGather, ReduceScatter, and point-to-point (SendRecv) operations. Appendix~\ref{appendix:collectives} shows that R2CC-Balance consistently maintains 85-89\% of normal throughput across all three operations.

\section{Other Related Work}

\smallskip\noindent\textbf{Collective communication optimization.} Prior work improves distributed ML communication by optimizing or automatically generating collective schedules~\cite{shah2023taccl,liu2024rethinking,devraj2025efficient}, and more recently by co-designing collectives with the underlying topology or fabric~\cite{kumar2025pccl}. These approaches target communication efficiency under normal or statically specified conditions, but do not address how to preserve an in-flight collective or adapt communication online after partial network failures. R2CC instead focuses on CCL-layer resilience under failures, combining online repair with post-failure schedule adaptation.

\smallskip\noindent\textbf{Network diagnostics and monitoring.}
There have been efforts to diagnose network issues in GPU clusters. System-level diagnosis and monitoring frameworks~\cite{liu2023hostping,liu2024hostmesh,liu2024r} isolate bottlenecks and localize likely fault sources using host- and network-level measurements in RDMA/RoCE clusters. A complementary line develops algorithmic telemetry mechanisms, including sketch-based monitoring systems~\cite{liu2016one, zhang2024octosketch}, for general and fine-grained network measurement. These works improve observability and fault localization, but they do not repair an in-flight collective or adapt collective schedules after partial inter-node failures.

\section{Conclusions}

In this paper, we introduce R2CC, a fault-tolerant collective communication library designed to sustain large-scale ML workloads through network failures without costly interruptions. By leveraging seamless connection migration, intelligent load balancing, and resilient communication algorithms, R2CC reduces failure-induced overheads.
Our experiments demonstrate that, in the presence of network failures, R2CC reduces training time overhead by up to 92\% and decreases inference overhead by up to 98\% compared to existing fault-tolerant methods.

\paragraph{Acknowledgments.} We thank the anonymous reviewers and our shepherd for their valuable feedback.
This work was supported in part by U.S. NSF grants IIS-2544434, CNS-2415754, and U.S. Army Research Laboratory Grant W911NF-25-2-0028.
This work used Bridges-2~\cite{brown2021bridges} at the Pittsburgh Supercomputing Center through allocation CIS251243 from the Advanced Cyberinfrastructure Coordination Ecosystem: Services \& Support (ACCESS) program~\cite{boerner2023access}, which is supported by U.S. NSF grants \#2138259, \#2138286, \#2138307, \#2137603, and \#2138296.

\bibliographystyle{plain}
\bibliography{reference}

\clearpage
\appendix
\section{Derivation and Approximation Gap for $Y = X$}\label{appendix:allreduce-math}

This appendix derives the closed-form optimal partition $Y^*$ for R2CC-AllReduce and the approximation-gap bound for the $Y = X$ rule stated in Section~\ref{sec:single-failure}, then tabulates the actual per-$(n, X)$ gap values. Notation follows the main text: $D$ is the per-node data size, $B$ the per-node bandwidth, $n \geq 2$ the number of nodes, $X \in (0, 1)$ the bandwidth loss fraction on the failed node, and $Y \in [0, 1]$ the fraction of data routed through the partial AllReduce. The no-failure ring AllReduce time is $\frac{2(n-1)}{n} \cdot \frac{D}{B}$.

\paragraph{Stage times.} Stage 1 runs two AllReduces concurrently. The global AllReduce spans all $n$ nodes on $(1-Y)D$ and is bottlenecked by the failed node's $(1-X)B$:
$$T_1(Y) = \frac{2(n-1)}{n} \cdot \frac{(1-Y)D}{(1-X)B}.$$
The partial AllReduce involves the $n-1$ healthy nodes on $YD$ using their reserved $XB$:
$$T_2(Y) = \frac{2(n-2)}{n-1} \cdot \frac{YD}{XB}.$$
Stage 2 broadcasts the $YD$ partition back to the failed node, limited by its $(1-X)B$ receive bandwidth:
$$T_3(Y) = \frac{YD}{(1-X)B}.$$
The total completion time is $T(Y) = \max(T_1(Y), T_2(Y)) + T_3(Y)$.

\paragraph{Optimal partition.} $T_1$ strictly decreases and $T_2$ strictly increases in $Y$, so $\max(T_1, T_2)$ is minimized at the crossover $T_1(Y^*) = T_2(Y^*)$, giving
$$Y^* = \frac{(n-1)^2 X}{n^2 - 2n + X}.$$
For $Y \in [0, Y^*]$, $T(Y) = T_1(Y) + T_3(Y) = \frac{D}{nB(1-X)}\bigl[(-n+2)Y + (2n-2)\bigr]$ is non-increasing since $-(n-2) \leq 0$ for $n \geq 2$. For $Y \in (Y^*, 1]$, $T(Y) = T_2(Y) + T_3(Y) = YD\bigl[\frac{2(n-2)}{(n-1)XB} + \frac{1}{(1-X)B}\bigr]$ is strictly increasing. Therefore $T(Y) \geq T(Y^*)$ for all $Y \in [0, 1]$, so $Y^*$ is optimal. At $n = 2$ the coefficients degenerate and $T(Y)$ is constant, so the algorithm reduces to R2CC-Balance; for $n \geq 3$ the minimum at $Y^*$ is strict, and R2CC-AllReduce utilizes the $XB$ healthy-node bandwidth that would otherwise sit idle during the bottlenecked global ring.

\paragraph{Gap bound for $Y = X$.} Since $Y = X < Y^*$ lies in the non-increasing region $[0, Y^*]$ of $T(Y)$, algebra gives the closed form
$$g(n, X) := \frac{T(X) - T(Y^*)}{T(Y^*)} = \frac{Y^* - X}{\frac{2(n-1)}{n-2} - Y^*}.$$
Because $\frac{2(n-1)}{n-2} \geq 2$ for $n \geq 3$ and $Y^* < 1$, the denominator exceeds $1$, so the gap is strictly less than $Y^* - X$. Using the identity $Y^* - X = \frac{X(1-X)}{n(n-2) + X}$ together with $X(1-X) \leq 1/4$:
$$g(n, X) < Y^* - X \leq \frac{1}{4n(n-2)}.$$
At $n = 2$ the bound is vacuous because R2CC-AllReduce degenerates to R2CC-Balance.

\begin{figure}[t]
    \centering
    \includegraphics[width=0.9\columnwidth]{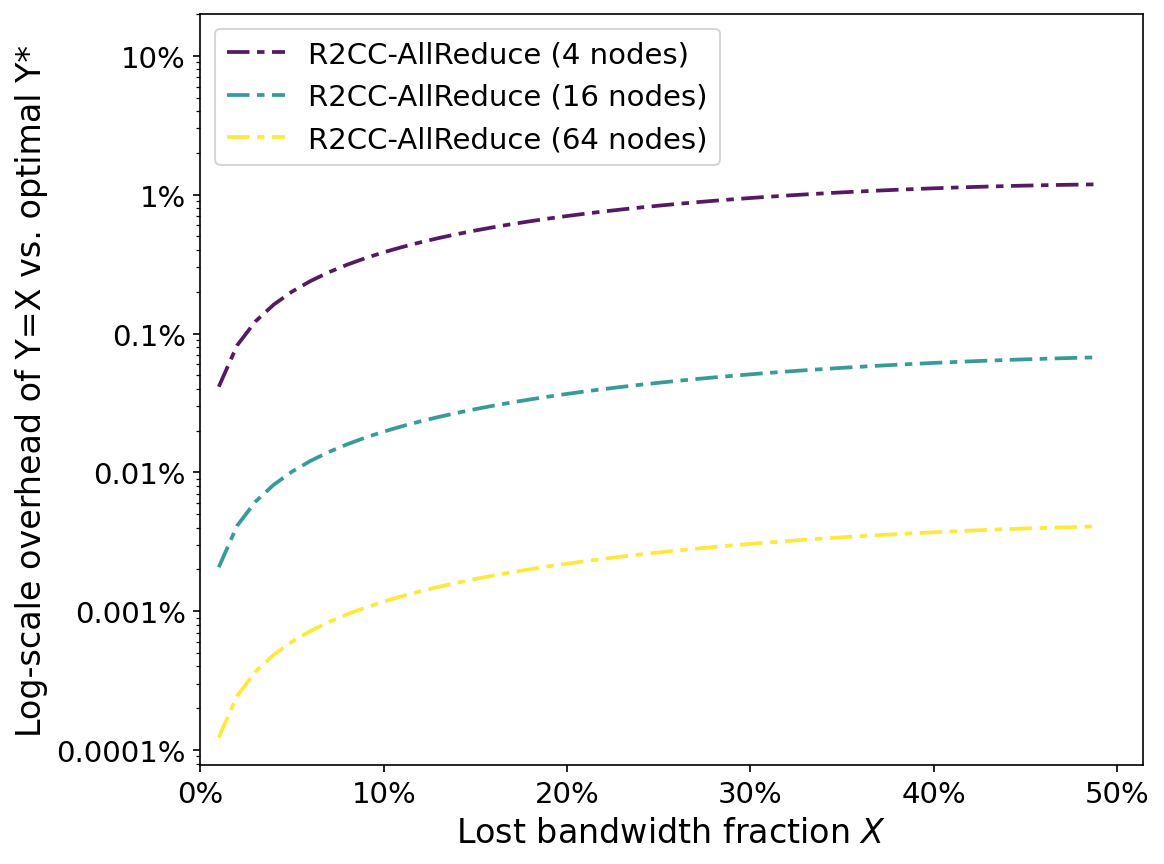}
    \tightcaption{Relative approximation gap $g(n, X) = (T(X) - T(Y^*))/T(Y^*)$ of the $Y = X$ rule vs.\ lost bandwidth fraction $X$, for $n \in \{4, 16, 64\}$. $y$-axis on log scale.}
    \label{fig:appendix-approx-gap}
\end{figure}

Table~\ref{tab:approx-gap-max} lists $g(n, X)$ at three representative loss fractions $X \in \{1/8, 1/4, 1/2\}$ and three node scales $n \in \{4, 16, 64\}$, matching the $n$ values plotted in Figure~\ref{fig:appendix-approx-gap}. Even at $n = 4$ and the worst-case $X = 1/2$, the gap is only $1.19\%$; at $n = 16$ it is below $0.07\%$ across all three $X$, and at $n = 64$ it drops below $0.005\%$. $Y = X$ is therefore a near-optimal choice across all practical configurations.

\begin{table}[h]
\centering
\small
\begin{tabular}{rccc}
\toprule
$n$ & $X = \tfrac{1}{8}$ & $X = \tfrac{1}{4}$ & $X = \tfrac{1}{2}$ \\
\midrule
4 & $0.470\%$ & $0.833\%$ & $1.190\%$ \\
16 & $0.02419\%$ & $0.04419\%$ & $0.06783\%$ \\
64 & $0.00145\%$ & $0.00265\%$ & $0.00411\%$ \\
\bottomrule
\end{tabular}
\caption{Relative approximation gap $g(n, X) = (T(X) - T(Y^*))/T(Y^*)$ at representative $(n, X)$ values.}
\label{tab:approx-gap-max}
\end{table}

Figure~\ref{fig:appendix-approx-gap} plots $g(n, X)$ as a continuous function of $X$ for $n \in \{4, 16, 64\}$. Each $4\times$ increase in $n$ shifts the curve down by roughly $1.2$ decades on the log $y$-axis, matching the $1/n^2$ asymptotic decay.

\section{Detailed Overhead Analysis}\label{appendix:overhead}

The overhead of using multiple NICs is confined to the I/O and host memory rather than to GPU memory or NVLink if using PCIe and QPI. When direct GPU–NIC access is available, R2CC registers the same GPU buffer with all NICs once at communicator creation and reuses this mapping for rerouted traffic, so no additional device memory is allocated. When a GPU–NIC pair cannot use direct access or cross NUMA, R2CC instead binds the collective to pinned host buffers and relies on CPU‑assisted copies, which likewise keep the GPU’s memory footprint unchanged. Prior work~\cite{liu2023hostping} shows that cross‑socket paths to a remote‑NUMA NIC can sustain more than half of the NIC’s line rate with only modest latency increase, which is sufficient to carry the fraction of traffic shifted off a failed NIC. When R2CC chooses PXN to avoid oversubscribed PCIe root complexes or CPU interconnects, the proxy GPU needs to allocate a small staging buffer on the order of tens of megabytes to relay segments to its local PCIe‑attached NIC, incurring one extra hop on NVLink but still negligible GPU‑memory overhead compared to model state.

\section{Detailed R2CC Failure Handling Scope}\label{app:failure-scope}

Table~\ref{tab:r2cc-failure-scope} summarizes R2CC's failure-handling scope, separating failures R2CC handles in-flight (fully or partially) from out-of-scope fundamentals that are left to higher-layer recovery.

\paragraph{Gray failures and link flapping.}\label{app:gray-failure}
R2CC detects failures through transport-level errors, such as work request completions that exceed the retry limit, and initiates recovery as described in \S\ref{sec:design-mig}. However, link flapping or packet loss masked by RDMA retransmissions does not produce such errors, leaving R2CC unaware of the degradation. Consequently, fault localization and HotRepair are not triggered, and collective operations continue to complete correctly, but at reduced throughput (Table~\ref{tab:r2cc-failure-scope}).
While detecting such gray failures is a limitation of the current design, R2CC-Balance (\S\ref{sec:single-failure}) already distributes traffic in proportion to per-NIC bandwidth. A possible extension would therefore be to use observed per-channel bandwidth to drive rebalancing, potentially mitigating gray failures that manifest as sustained throughput degradation.

\begin{table*}[!t]
\centering
\begin{tabular}{p{0.30\textwidth}p{0.12\textwidth}p{0.50\textwidth}}
\toprule
\textbf{Failure Type} & \textbf{Support} & \textbf{When Supported (Boundary)} \\
\midrule

\multicolumn{3}{l}{\textit{In-flight inter-node communication failures (R2CC’s primary scope)}} \\
\midrule
NIC hardware / port failure (incl. NIC--ToR) & Yes &
Node/process stays alive, and the node still has at least one remaining healthy inter-node NIC/path. \\
\addlinespace

Inter-node link / cable / ToR port down (single rail/link) & Yes &
An alternate inter-node path exists (e.g., another rail/NIC); not a full network partition. \\
\addlinespace

RDMA transport / QP-level failure (error CQE, QP error, WQE flush) & Yes &
Failure is confined to a subset of connections while endpoints remain alive and an alternate NIC/path exists. \\
\addlinespace

Link flapping (up$\rightarrow$down$\rightarrow$up) & Partial &
Only when flapping causes an in-flight transport failure (e.g. timeout) visible to the collective; pure throughput jitter is not handled. \\
\addlinespace

Multi-hop fabric fault (Inter-switch link / spine switch failure) & Partial &
Usually masked by fabric rerouting; if it surfaces as a transport error, handled via backup connections. \\
\addlinespace

CRC error (packet corruption) & Partial &
Only when CRC errors escalate into an in-flight transport failure; otherwise typically no recovery action is triggered. \\
\addlinespace

NIC driver issue & Partial &
Supported only if it does not crash the OS/process and still leaves an alternate NIC/path usable. \\
\addlinespace

NIC firmware issue & Partial &
Supported only if it degrades a subset of NICs while the node/process remains alive and other NICs/paths remain usable. \\
\addlinespace

PCIe failure (NIC unreachable / disappears) & Partial &
Supported when PCIe issues take down only a subset of NICs while other NICs remain usable; system-wide I/O failure is out of scope. \\
\addlinespace

GPU$\leftrightarrow$NIC direct path unavailable (GPUDirect / PCIe P2P degraded) & Partial &
Supported only if communication can continue via other usable inter-node NIC/path (potentially with degraded performance). \\

\midrule
\multicolumn{3}{l}{\textit{Out-of-scope fundamentals}} \\
\midrule

NVLink / NVSwitch failure (intra-node fabric) & No &
Not supported. \\
\addlinespace

Network partition / switch-wide outage (no alternate path) & No &
Not supported. \\
\addlinespace

Node/OS/process crash / GPU failure & No &
Not supported. \\
\addlinespace

Cross-rail mistaken wiring & No &
Out of scope; assumes the job can initialize and run normally. \\
\bottomrule
\end{tabular}
\caption{\textbf{R2CC failure model and scope.} Support means the system can keep an ongoing collective running without communicator re-initialization or job restart, under the stated boundary conditions.}
\label{tab:r2cc-failure-scope}
\end{table*}

\section{Optimize Scheduling Under Multi-Failure}
\label{sec:multi-failure}
Large-scale distributed GPU clusters can face concurrent failures~\cite{Revisiting_Reliability}.
Extending single-failure handling to support multi-failure resilience introduces two formidable challenges:
(1) Topological Fragmentation: In rail-optimized topologies, asymmetric failures on adjacent nodes (e.g., Node A loses NIC 1, while Node B loses NIC 2) disrupt the alignment of parallel rails, rendering simple local load balancing insufficient.
(2) Bandwidth Spectrum: Multiple failures create a spectrum of available bandwidth capacities across nodes, rather than a binary healthy/faulty distinction.
In this section, we present how R2CC generalizes its scheduling primitives to address these complexities through \textit{Topology-Aware Logical Re-ranking} and \textit{Recursive AllReduce Decomposition}.

\begin{figure}
    \centering
    \includegraphics[width=0.7\columnwidth]{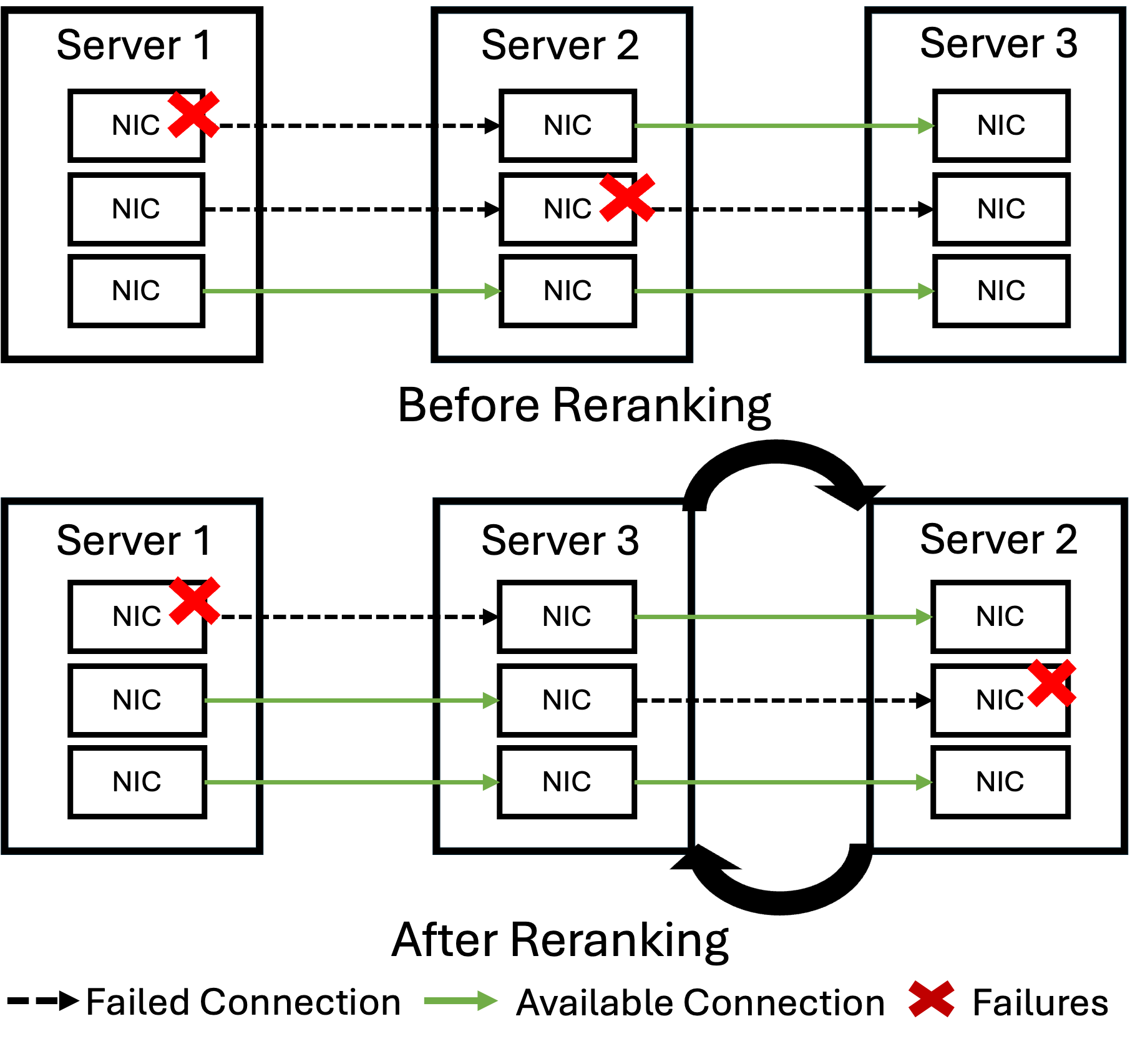}
    \vspace{-5pt}
    \tightcaption{\textbf{Topology-aware logical re-ranking.}}
    \label{fig:Bandwidth Bridge}
\end{figure}

\paragraph{Topology-Aware Logical Re-ranking.} NCCL typically fixes ring/tree order by physical locality or static configuration. When adjacent nodes experience disjoint NIC failures (e.g., node $u$ loses rail $r$ while neighbor $v$ loses rail $r'$), their shared bandwidth drops to the narrow intersection of surviving rails, and local load balancing can no longer compensate.

We observe that most collective algorithms are symmetric and agnostic to node ordering, enabling safe reordering. It performs topology-aware logical re-ranking (Figure~\ref{fig:Bandwidth Bridge}): pairs of neighbors whose rail overlap falls below a bandwidth threshold are separated by inserting “bridge’’ nodes with broader rail connectivity. This targeted repair modifies only the problematic edges rather than rebuilding the entire ring, preserving most RDMA connections.

\paragraph{Recursive R2CC-AllReduce.} The R2CC-AllReduce strategy in Section~\ref{sec:single-failure} isolates a single bottleneck node and reduces its communication load, assuming the remaining nodes are homogeneous. Under concurrent failures, however, clusters develop a bandwidth spectrum, and treating all non-bottleneck nodes as a single group forces faster nodes to operate at the speed of moderately degraded ones, leaving substantial bandwidth unused.

R2CC addresses this by generalizing dual-ring decomposition into a recursive scheduling strategy. The system first forms a global ring running at the slowest node’s rate. It then isolates that node and builds a faster sub-ring from the remaining participants. If bandwidth variance persists, R2CC recursively peels off further sub-rings, each handling a data chunk proportional to the incremental bandwidth of its members. Logical re-ranking is applied at every level to avoid rail mismatches introduced by skipping slower nodes.

R2CC executes the reduction phases of all rings in parallel to saturate available bandwidth. Broadcast phases impose dependencies—slower nodes must wait for partial results from faster sub-rings—but bandwidth asymmetry often allows faster nodes to finish their smaller broadcasts concurrently with slower nodes' ongoing operations. To formalize these decisions, R2CC extends NCCL’s performance model ($\alpha$–$\beta$) to evaluate expected completion time at each recursion depth, incorporating per-node bandwidth and overlap. The planner then selects among standard Ring/Tree, R2CC-Balance, or recursive R2CC-AllReduce to best navigate latency–throughput trade-offs.

\begin{figure}[t]
    \centering
    \includegraphics[width=1\linewidth]{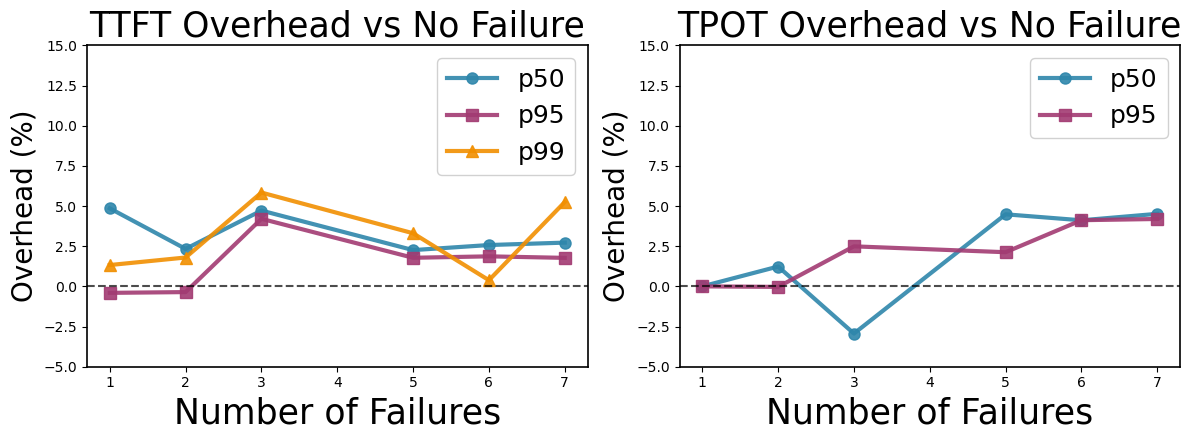}
    \caption{p50/p95 TTFT/TPOT overhead vs number of failures under multiple NIC failures for serving with pipeline parallelism}
    \label{fig:multi_failure}
\end{figure}

\paragraph{Severe single node and rail failures.} By the model of \S\ref{sec:single-failure}, losing 2 or 3 NICs on the same node ($X=1/4, 3/8$) slows each AllReduce by 17--22\% and 30--40\% with R2CC-AllReduce on 4--128 nodes, and by 33\% and 60\% with R2CC-Balance (table below). A ToR switch failure that takes out a full rail removes the same NIC from every node, leaving no spare bandwidth for R2CC-AllReduce, so R2CC-Balance operates at 14.3\% overhead for one of eight rails. End-to-end overheads are much smaller: on our testbed, two simultaneous NIC failures on one node add at most 1.01\% training overhead (\S\ref{exp:training}), and up to seven failed NICs on one node keep inference within about 5\% (Figure~\ref{fig:multi_failure}).

\begin{center}
\small
\setlength{\tabcolsep}{4pt}
\begin{tabular}{lcccccc}
\toprule
Failure & $X$ & Balance & $n{=}4$ & $16$ & $64$ & $128$ \\
\midrule
1 NIC, 1 node & $1/8$ & 14.3 & 9.5 & 7.6 & 7.3 & 7.2 \\
2 NICs, 1 node & $1/4$ & 33.3 & 22.2 & 17.8 & 16.9 & 16.8 \\
3 NICs, 1 node & $3/8$ & 60.0 & 40.0 & 32.0 & 30.5 & 30.2 \\
Full rail & $1/8$ & 14.3 & \multicolumn{4}{c}{no gain (same as Balance)} \\
\bottomrule
\end{tabular}
\end{center}

\section{R2CC Bridge-Based Re-ranking}
\label{app:bridge-based}

Algorithm~\ref{alg:reranking} inserts bridge nodes with broader rail connectivity between ring neighbors whose surviving rail overlap falls below the global bandwidth threshold, repairing only the problematic edges while leaving the rest of the ring intact.

\begin{algorithm}[t]
\small
\caption{R2CC Bridge-Based Re-ranking}
\label{alg:reranking}
\textbf{Input:} Logical Ring $R$, Rail Sets $S$ \\
\textbf{Output:} Optimized Ring $R'$
\begin{algorithmic}[1]
\STATE \textbf{Initialize} $R' \leftarrow R$, $Candidates \leftarrow \emptyset$
\STATE $B_{global} \leftarrow \min_{n \in R} |S_n|$
\FOR{$i \leftarrow 0$ \textbf{to} $|R| - 1$}
    \STATE $u \leftarrow R[i]$, $v \leftarrow R[(i+1) \pmod{|R|}]$
    \IF{$|S_u \cap S_v| < B_{global}$}
        \STATE $Candidates.\text{add}((u, v))$
    \ENDIF
\ENDFOR
\STATE \textbf{Sort} $Candidates$ by severity (gap size) descending
\FOR{each pair $(u, v)$ in $Candidates$}
    \STATE $BestBridge \leftarrow \text{None}$
    \FOR{each node $w \in R' \setminus \{u, v\}$}
        \STATE $x \leftarrow \text{PrevNode}(w), y \leftarrow \text{NextNode}(w)$
        \STATE $NewCap \leftarrow \min( |S_u \cap S_w|, |S_w \cap S_v| )$
        \STATE $RemovalCap \leftarrow |S_x \cap S_y|$
        \IF{$NewCap \ge B_{global}$ \textbf{and} $RemovalCap \ge B_{global}$}
             \STATE $BestBridge \leftarrow w$
             \STATE \textbf{break}
        \ENDIF
    \ENDFOR
    \IF{$BestBridge \neq \text{None}$}
        \STATE $R'$.Relocate($BestBridge$, between $u, v$)
    \ENDIF
\ENDFOR
\STATE \textbf{return} $R'$
\end{algorithmic}
\end{algorithm}

\section{Additional Evaluation Results}\label{appendix:collectives}

\paragraph{R2CC Inference Performance Under Multiple Failures.}\label{multi-failure-inference} We further evaluate R2CC's resilience under multiple concurrent failures on our two-node testbed. Unlike training workloads that may scale to hundreds of nodes, inference deployments typically use fewer nodes per model instance, reflecting production requirements for latency and cost efficiency. Our multi-failure experiments use vLLM engine to serve a 405B model with TP=8 and PP=2 configuration. We inject varying numbers of failures to reduce available NICs on a single node and feed requests at QPS=0.1, ensuring the system operates within its capacity to measure steady-state overhead. As shown in Figure \ref{fig:multi_failure}, both TTFT and TPOT overheads remain within 0-5\%, exhibiting minimal variance due to system underlying overhead, even as multiple network device failures reduce majority bandwidth. This demonstrates that R2CC fully exploits the ample bandwidth headroom in inference workloads: even under more severe and numerous failures on network devices, the remaining links still suffice to allow the system to continue serving with overhead close to zero.

\begin{figure}
    \centering
    \begin{subfigure}[b]{0.48\columnwidth}
        \includegraphics[width=\textwidth]{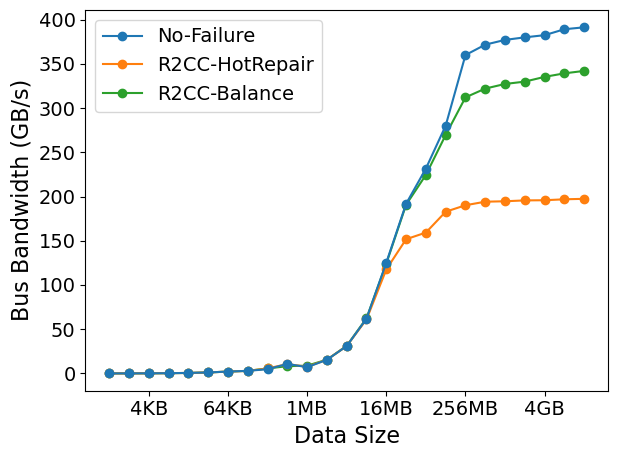}
        \caption{AllGather}
        \label{fig:allgather-h100}
    \end{subfigure}

    \vskip\baselineskip

    \begin{subfigure}[b]{0.48\columnwidth}
        \includegraphics[width=\textwidth]{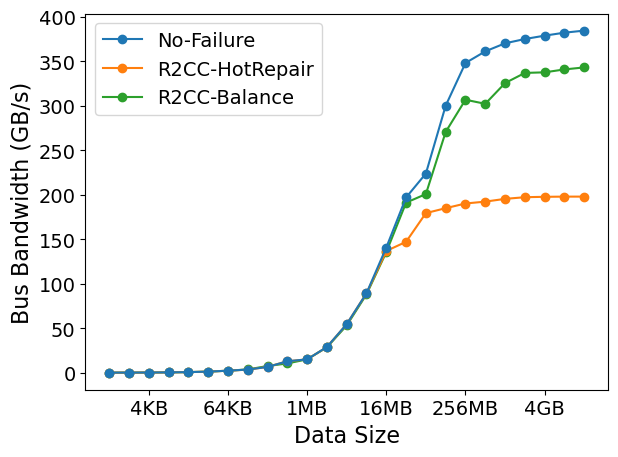}
        \caption{ReduceScatter}
        \label{fig:reducescatter-h100}
    \end{subfigure}
    \hfill
    \begin{subfigure}[b]{0.48\columnwidth}
        \includegraphics[width=\textwidth]{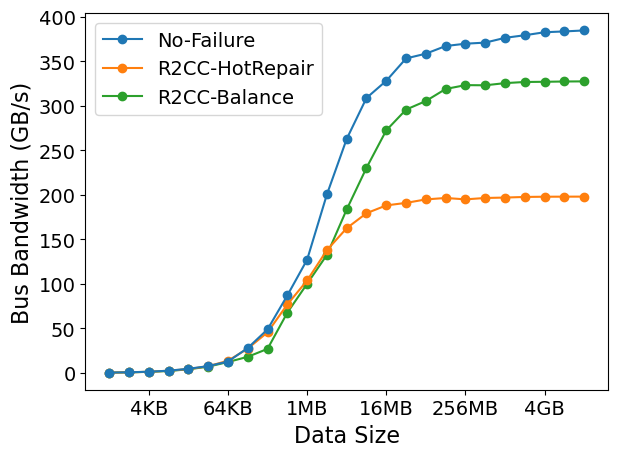}
        \caption{SendRecv}
        \label{fig:sendrecv-h100}
    \end{subfigure}

    \tightcaption{H100 NCCL-Test Benchmark.}
    \label{appendix:fig:nccl-test-allreduce}
\end{figure}

\paragraph{Other Collective and Point-to-Point Operations.}
Beyond AllReduce, we evaluate R2CC-Balance on AllGather, ReduceScatter, and point-to-point (SendRecv) operations. Figure~\ref{appendix:fig:nccl-test-allreduce} shows that R2CC-Balance consistently maintains 83--90\% of normal throughput across these three operations for large messages, while the R2CC-HotRepair approach suffers approximately 50\% throughput loss due to its single-NIC bottleneck.

These results demonstrate that R2CC's load-balancing mechanism generalizes effectively beyond AllReduce, providing consistent fault tolerance for the full range of communication primitives used in modern distributed ML systems.

\begin{figure}
    \centering
    \includegraphics[width=0.7\columnwidth]{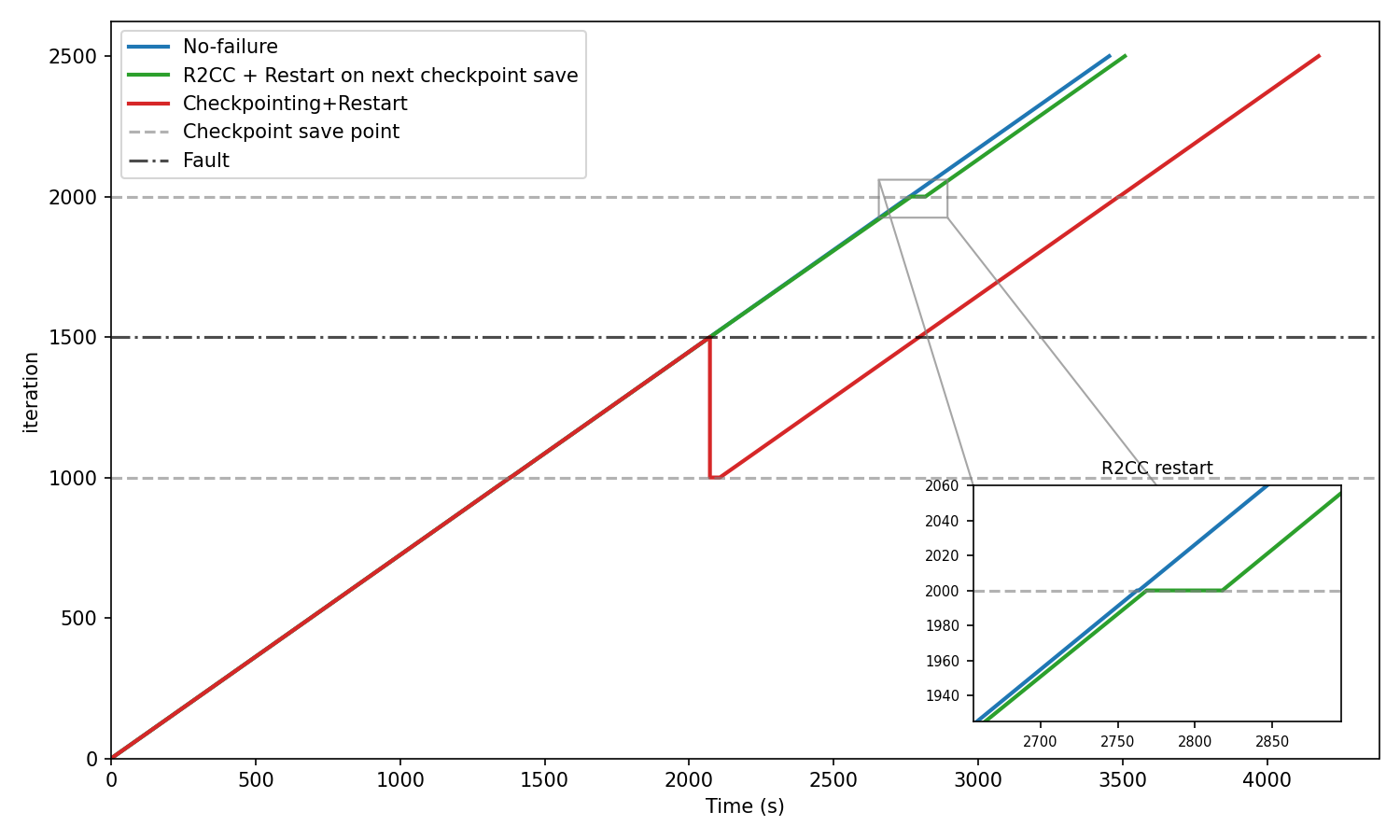}
    \tightcaption{\textbf{Checkpoint and restore in a training workload}}
    \label{fig:checkpoint-and-restore}
\end{figure}
\section{Checkpoint and restore}\label{appendix:checkpoint}
Figure \ref{fig:checkpoint-and-restore} shows a training run with a NIC failure injected mid-interval between two checkpoints. The naive baseline aborts at the failure and rolls back to the previous checkpoint, discarding the iterations since the last save; R2CC keeps training through to the next scheduled checkpoint, saves state, and only then restarts to recover the NIC, resuming without losing progress.

\section{Training Quality Under a NIC Failure}\label{app:quality}
\paragraph{Setup.} We train GPT-2 (124M)~\cite{radford2019language} on WikiText-103~\cite{merity2016pointer} with PyTorch DDP for 1,000 updates on three CloudLab servers, each with two V100S GPUs and three NIC ports. After update 400, we install a drop rule on one NIC port of one server for the rest of the run, leaving the other NICs up. Runs are deterministic, and each failure run is paired with a failure-free run of the same seed (three seeds per policy).

\begin{figure}[t]
    \centering
    \begin{subfigure}[b]{0.48\columnwidth}
        \includegraphics[width=\textwidth]{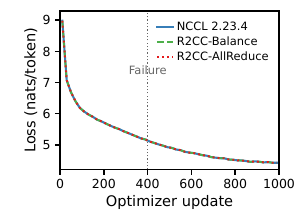}
        \caption{Training loss trajectory.}
        \label{fig:loss-traj}
    \end{subfigure}
    \hfill
    \begin{subfigure}[b]{0.48\columnwidth}
        \includegraphics[width=\textwidth]{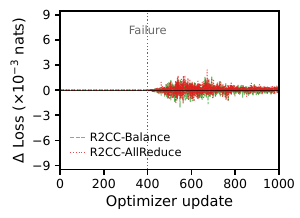}
        \caption{Training loss deviation.}
        \label{fig:loss-dev}
    \end{subfigure}
    \tightcaption{Training loss under a persistent NIC-port failure (GPT-2 124M, WikiText-103, 3 seeds).}
    \label{fig:quality}
\end{figure}

\paragraph{Results.} Without failures, R2CC is bit-identical to the upstream NCCL 2.23.4 it forks from: the same losses and the same hashes of reduced gradients and parameters at every checked update. Under the failure, R2CC completes the interrupted AllReduce on the backup connection and continues under R2CC-Balance or R2CC-AllReduce without restart or rollback. HotRepair itself is lossless: the repaired collective at update 401 produces a bit-identical reduced gradient. Differences begin only at update 402, where R2CC-Balance and R2CC-AllReduce redistribute data across the remaining NICs and thereby change the order in which gradients are summed; floating-point addition is not associative, so a different order rounds differently. To verify that reordering is the only effect, we re-sum the recorded gradient inputs in double precision as an exact reference: both NCCL's and R2CC's results stay within a relative error of $5.8\times10^{-8}$ of it, i.e., ordinary FP32 rounding. Figure~\ref{fig:quality} shows the paired trajectories: the per-update loss deviation stays within $2.4\times10^{-3}$ nats/token with no consistent sign, and final test perplexity rises by at most 0.023\% (Table~\ref{tab:quality}).

\begin{table}[h]
\centering
\small
\begin{tabular}{lrr}
\toprule
Condition & Test PPL & Max.\ paired $\Delta$ \\
\midrule
NCCL 2.23.4, no failure & 76.085 & -- \\
R2CC, no failure & 76.085 & $0.000\%$ \\
R2CC-Balance, failure & 76.085 & $+0.008\%$ \\
R2CC-AllReduce, failure & 76.086 & $+0.023\%$ \\
\bottomrule
\end{tabular}
\caption{Final test perplexity (mean of three seeds). $\Delta$ is the worst-case increase over the NCCL baseline run with the same seed.}
\label{tab:quality}
\end{table}

\section{Discussion}\label{app:discussion}
\paragraph{Multi-plane fabrics and direct-attached NICs.} Multi-plane fabrics connect each NIC to several independent switch planes, so a link or switch failure in one plane is masked at reduced bandwidth. The affected NIC then runs slower than its peers, and R2CC's schedule optimization could further rebalance traffic toward the other NICs once the fabric exposes this reduced bandwidth. A fault in the NIC itself, such as a hardware, firmware, or PCIe error, still takes down all of its planes, and R2CC handles it by failing over to another NIC. In GPU servers that attach each NIC directly to its GPU, such as HGX B300, a GPU has no direct PCIe path to the NICs of other GPUs, but when its NIC fails, R2CC can still relay its traffic over NVLink to GPUs with healthy NICs through PXN.

\paragraph{SHARP.} NVLink SHARP performs reductions inside the NVSwitch without traversing NICs, so it is complementary to R2CC, which protects inter-node transfers. InfiniBand SHARP instead keeps partial aggregates in switch memory, so a NIC failure during an in-network reduction cannot be repaired by migrating connections, and R2CC does not currently handle it. Supporting SHARP, e.g., by falling back to the host-based path or redistributing data across the aggregation trees of the surviving rails, is left to future work.

\paragraph{CUDA Graphs.} R2CC is compatible with CUDA Graphs. Hot repair only changes which connection the CPU proxy uses, and NCCL re-drives the proxy on every replay, so a graph captured before a failure keeps replaying over the backup NIC with its captured kernels and buffer addresses unchanged. The schedule optimizations of \S\ref{sec:single-failure} are decided when a collective is enqueued, so they take effect once the framework re-captures the graph, e.g., upon R2CC's degradation signal (\S\ref{sec:signal-restore}). A large graph does not need to be re-captured as a whole, since only the child graph that holds the collective has to be re-captured and swapped in. Re-capture only records operations without executing them, so capturing and instantiating a graph containing the collective takes about one to two milliseconds, and the next replay already runs the new schedule. R2CC-AllReduce must create its sub-communicators outside capture, so it needs one eager collective after repair, which adds a one-time setup of about 0.3\,s to a regular iteration. Table~\ref{tab:cuda-graphs} shows these behaviors on a three-server CloudLab testbed.

\begin{table}[t]
\centering
\small
\begin{tabular}{lrr}
\toprule
 & Time & Re-capture \\
Replayed graph (4\,GiB AllReduce) & / iter & overhead$^{*}$ \\
\midrule
Pre-failure graph, healthy & 2.03\,s & -- \\
Pre-failure graph, HotRepair & 4.07\,s & -- \\
Re-captured as R2CC-Balance & 3.04\,s & 1.7\,ms \\
Re-captured as R2CC-AllReduce & 2.79\,s & 0.33\,s \\
\bottomrule
\end{tabular}
\caption{CUDA Graphs on three CloudLab servers, each with two V100S GPUs and three NIC ports limited to 10\,Gbps. $^{*}$Re-capture overhead is the time of the iteration that performs the re-capture, including the eager collective that R2CC-AllReduce needs to create its sub-communicators, minus the time of an iteration replayed from the new graph.}
\label{tab:cuda-graphs}
\end{table}

\section{Artifact Appendix}\label{app:artifact}

\subsection*{Abstract}

The artifact contains R2CC, a fault-tolerant collective communication library implemented as a drop-in replacement for NCCL 2.23.4, and an experiment suite that injects NIC-port failures on a multi-NIC GPU cluster and checks HotRepair, post-failure schedule optimization, and training behavior. Existing ML frameworks use R2CC without modification by linking against it instead of upstream NCCL.

\subsection*{Scope}

Since the H100 testbed in \S\ref{sec:setup} is a shared cluster that cannot be held for long, the artifact is evaluated on CloudLab\cite{Duplyakin+:ATC19} r7525 servers. It validates the following claims:
(1) R2CC recovers in-flight collectives from a NIC-port failure quickly, losslessly, and correctly instead of aborting the job (\S\ref{sec:design-mig});
(2) the R2CC-Balance and R2CC-AllReduce scheduling optimizations follow the trends reported in \S\ref{sec:microbenchmark}, and their measured performance is within 1\% of what the model in \S\ref{sec:single-failure} predicts for this testbed;
(3) R2CC is compatible with CUDA Graphs (Appendix~\ref{app:discussion});
(4) training quality is preserved under a NIC failure (Appendix~\ref{app:quality}).
Collective results are checked for correctness on both healthy and failed networks.

\subsection*{Contents}

The repository contains the R2CC source code on top of NCCL 2.23.4 (\texttt{src/}, with the NCCL build files) and the CloudLab experiment suite in \texttt{examples/cloudlab\_r7525/}. The suite has nine scripts: tests 01--02 check in-flight recovery (about 40\,s each), tests 03--07 check the scheduling optimizations with nccl-tests (about 16\,min in total), test 08 checks CUDA Graphs (about 1.5\,min), and test 09 trains GPT-2 with PyTorch DDP through a NIC failure (about 45\,min).

\subsection*{Hosting}

The artifact is available at \url{https://github.com/r2cc-project/R2CC/tree/artifact_evaluation}, branch \texttt{artifact\_evaluation}, commit \texttt{e93425b}.

\subsection*{Requirements}
Evaluating R2CC requires at least two GPU servers, each with multiple NICs. We use CloudLab r7525 servers with BlueField-2 SmartNICs, available as the r7525-bf2 node type at Clemson, and provide a public CloudLab profile and a pre-built image. The provided experiments use three such servers, and NIC-port failures are injected through their BlueField-2 SmartNICs.

\subsection*{Setup and Reproduction}

The README in the \texttt{artifact\_evaluation} branch gives the complete setup and reproduction guide, and \texttt{examples/cloudlab\_r7525/r7525\_setup.md} describes the node setup. Running all nine tests takes about one hour.

\end{document}